\documentclass[Journal, InsideFigs,, DoubleSpace, NoLineNumbers, letterpaper]{ascelike-new}

\usepackage[utf8]{inputenc}
\usepackage[T1]{fontenc}
\usepackage{lmodern}
\usepackage{graphicx}
\usepackage{xcolor}
\usepackage{caption}
\usepackage{subcaption}
\usepackage{amsmath}
\usepackage{amssymb}
\usepackage{newtxtext,newtxmath}
\usepackage[colorlinks=true,citecolor=red,linkcolor=black]{hyperref}
\usepackage{tabularx}
\usepackage{adjustbox}
\usepackage[authoryear]{natbib}
\usepackage{multirow}
\usepackage{booktabs}
\usepackage{pdflscape}
\usepackage{fancyhdr}
\usepackage{everypage}
\usepackage{setspace}
\usepackage{rotating}
\usepackage{algorithm}
\usepackage{algpseudocode}



\newcommand{\Lpagenumber}{\ifdim\textwidth=\linewidth\else\bgroup
  \dimendef\margin=0
  \ifodd\value{page}\margin=\oddsidemargin
  \else\margin=\evensidemargin
  \fi
  \raisebox{\dimexpr -\topmargin-\headheight-\headsep-0.5\linewidth}[0pt][0pt]{
    \rlap{\hspace{\dimexpr \margin+\textheight+\footskip}
    \llap{\rotatebox{90}{\thepage}}}}
\egroup\fi}
\AddEverypageHook{\Lpagenumber}

\NameTag{Rostami, \today}

\begin{document}

\title{Inspect Any Crack: An Automated End-to-End Crack Segmentation and Severity Assessment Framework}

\author[1]{Ghodsiyeh Rostami, MSc}
\author[2]{Po-Han Chen, Ph.D.}
\author[3]{Mahdi S. Hosseini, Ph.D.}

\begingroup
\singlespacing
\affil[1]{PhD Candidate, Department of Building, Civil, and Environmental Engineering,
          Concordia University, Montreal, Canada, H3G 1M8. Email: rose.rostami@concordia.ca}
\affil[2]{Professor, Department of Building, Civil, and Environmental Engineering,
          Concordia University, Montreal, Canada, H3G 1M8. Email: pohan.chen@concordia.ca
          (Corresponding author)}
\affil[3]{Assistant Professor, Department of Computer Science and Software Engineering,
          Concordia University, Montreal, Canada, H3G 1M8.
          Email: mahdi.hosseini@concordia.ca}
\maketitle
\endgroup

\begin{abstract}
Image-based crack analysis is essential for automated infrastructure inspection to reduce the cost, subjectivity, and safety risks of manual surveys. Despite advances in visual crack detection, methods that translate predictions into physically meaningful and engineering-interpretable crack condition information remain limited. To bridge this gap, this study proposes Inspect Any Crack (IAC), an innovative end-to-end crack recognition and condition assessment framework that automatically derives physical geometric measurements and continuous severity scores from concrete-crack imagery. At its core, IAC introduces SACv2-Tiny, an efficient crack segmentation model adapted from a vision foundation model. Trained on the challenging OmniCrack30K dataset, SACv2-Tiny achieved a $62.49\%$ F1-score and $45.45\%$ IoU on the test set and a mean zero-shot F1-score of $69.89\%$ across three unseen datasets, demonstrating robust segmentation that enables downstream analysis with limited post-processing. IAC further processes the predicted masks to quantify physical crack width and affected area using ground-sampling-distance calibration. Validation at 41 caliper-referenced locations yielded a mean relative width error of $7.33\%$. Beyond geometric quantification, IAC introduces a novel continuous crack-condition metric, the Crack Severity Index (CSI), bounded on $[0,100]$, which combines maximum crack width and affected-area ratio into a unified, standards-informed severity measure. Through these developments, IAC advances crack inspection beyond pixel-level recognition toward interpretable and automated infrastructure condition assessment.

\end{abstract}

\section{Practical Applications}
Inspecting concrete infrastructure such as bridges and buildings is essential for timely maintenance. Meanwhile, manual surveys and contact-based crack measurements are slow, subjective, costly, and may
expose inspectors to hazards. Automated camera- and drone-based inspection can reduce these burdens, yet most
developed image-based methods stop after locating crack pixels and do not provide the physical measurements
needed for maintenance decisions. This study introduces Inspect Any Crack (IAC), a framework that
automatically segments cracks, estimates their width and affected area, converts measurements
into physical units, and produces a continuous severity score that can be aligned with inspection
standards. Tests on diverse image datasets showed robust crack detection. At 41 measured locations,
estimated widths differed from digital caliper measurements by 7.33\% on average when images met
the recommended resolution. 
By combining maximum crack width with affected area, the severity score distinguishes
localized cracks from distributed cracking. The framework can support consistent, traceable, and
safer condition assessment using handheld cameras, mobile robots, or drones.

\section{Introduction} \label{sec:introduction}

Aging bridges and buildings pose increasing safety and serviceability risks
\citep{xiangCracksegmentationAlgorithmFusing2023,munawarImageBasedCrackDetection2021}.
Cracks are common early indicators of distress and can accelerate reinforcement corrosion and
spalling if untreated \citep{munawarImageBasedCrackDetection2021}. Routine inspection enables timely
maintenance, however, manual surveys and contact measurements are labor-intensive, costly, subjective,
and potentially hazardous \citep{halderRobotsInspectionMonitoring2023,
agnisarmanSurveyAutomationenabledHumanintheloop2019}.
Automated visual inspection uses UAVs and advanced imaging to improve the safety and efficiency of
remote surveys \citep{pengUAVbasedMachineVision2021,zhangFullyAutomatedUnmanned2022}.
In recent years, advancements in deep semantic segmentation have progressed pixel-level crack localization
\citep{guoSurfaceDefectDetection2024,iraniparastSurfaceConcreteCracks2023}, where methods based on vision foundation
models such as SAC \citep{rostami2026segment}, CrackSAM \citep{geFinetuningVisionFoundation2024} have improved crack segmentation accuracy and zero-shot generalization \citep{guoSegmentAnythingModelbased2025,chenSECrackSegHighAccuracyCrack2025}. 
However, a segmentation mask alone does not provide an engineering condition assessment, where maintenance decisions require physical crack measurements and interpretable severity indicators.

In structural inspection practice, crack width is a key geometric indicator used to characterize
crack condition.  ACI~224R-01 and
AASHTO classify widths below $0.2$\,mm as hairline, $0.2$-$0.5$\,mm as fine,
$0.5$-$1.0$\,mm as medium, and above $1.0$\,mm as wide. In Canada, the Ontario
Structure Inspection Manual (OSIM; MTO, 2008) uses corresponding thresholds of
$0.1$, $0.3$, and $1.0$\,mm, designating the $0.1$-$0.3$\,mm class as narrow.
Segmentation masks, however, still require inspectors to measure physical widths with gauges or
comparator cards, a subjective and time-consuming follow-up \citep{dengReviewComputerVisionbased2022b}.
Few studies have combined segmentation with geometric measurement
\citep{yangAutomaticPixelLevelCrack2018,kangHybridPixellevelConcrete2020,
dingCrackDetectionQuantification2023,liAutomatedQuantificationCrack2024}, where outputs often
remain in pixels or depend on a fixed imaging setup.
Limited studies have translated detected cracks into condition or severity assessments
\citet{wangAutomatedCrackSeverity2021} classified track-slab crack images into three severity levels, while the method lacks physical width measurements to enable alignment with different inspection standards.
Similarly, \citet{shiDeepLearningAutomated2025} classified concrete surface images into four dataset-defined condition categories without producing calibrated crack geometry.
Thus, automated concrete inspection still lacks an integrated image-based method that converts per-image calibrated crack geometry into a unified, traceable, and continuous severity score aligned with inspection standards.

To address this gap, this study proposes Inspect Any Crack (IAC), an end-to-end crack recognition and condition assessment framework that converts field images into calibrated crack geometry and continuous severity scores. As illustrated in Figure~\ref{fig:pipeline}, the framework comprises three stages: (1) crack segmentation, (2) geometric analysis and physical calibration, and (3) severity assessment. For segmentation, Segment Any Crackv2-Tiny (SACv2-Tiny) is developed through efficient adaptation and systematic evaluation of SAM2 variants, with SAM2-Tiny selected for its balance between accuracy and computational efficiency \citep{rostami2026segment,raviSAM2Segment2024}. The predicted masks are subsequently processed to estimate crack width and affected area using distance-based geometric analysis, with reference-object or ground-sampling-distance (GSD) calibration converting pixel measurements to physical units. Finally, a continuous Crack Severity Index (CSI) combines maximum crack width and affected-area ratio on a bounded $[0,100]$ scale, supporting both severity ranking and standards-informed classification. SACv2-Tiny is evaluated on OmniCrack30k and three zero-shot datasets. Physical measurements are validated against caliper-referenced concrete cracks, and the CSI is evaluated using measured and synthetic scenarios.
The contributions of the study are as follows:
\vspace{-10pt}
\begin{itemize}
    \item An end-to-end automated inspection framework, Inspect Any Crack (IAC), is developed to integrate crack segmentation, calibrated geometric measurement, and standards-informed severity assessment, providing a complete pipeline from field imagery to a quantitative crack severity score.

    \item A robust and computationally efficient crack segmentation model, Segment
          Any Crackv2-Tiny (SACv2-Tiny), is developed based on SAM2 vision foundation model, which demonstrates strong zero-shot
          generalization across unseen datasets.

\item A calibrated crack geometry module is developed and experimentally validated against caliper-referenced measurements, with four width estimation algorithms systematically compared to identify the most accurate estimator.

    \item A novel Crack Severity Index (CSI) is proposed for surface concrete crack condition ranking, integrating maximum crack width and affected-area measures
          into a continuous, bounded score on $[0,100]$ whose reference width and
          decision thresholds can be aligned with requirements of crack inspection manuals (i.e.,
          ACI~224R-01, AASHTO, and OSIM criteria)
\end{itemize}

The remainder of the paper is organized as follows. Section~\ref{sec:related_work} reviews related
work, Section~\ref{sec:methodology} describes the framework and evaluation, Section~\ref{sec:results}
presents the results, and Section~\ref{sec:conclusion} summarizes the findings and future work.

\section{Related Work} \label{sec:related_work}

Crack segmentation has progressed from CNN encoder-decoder architectures such as U-Net
\citep{ronnebergerUNetConvolutionalNetworks2015} and DeepLabv3+
\citep{chenEncoderDecoderAtrousSeparable2018a} to transformer- and state-space-based models, including
CrackFormer \citep{liuCrackFormerTransformerNetwork2021}, SegFormer
\citep{xieSegFormerSimpleEfficient2021}, and MambaCrackNet
\citep{hanEnhancingPixellevelCrack2024}. More recently, Vision Foundation Models (VFMs) have been adapted to crack segmentation to improve transferability beyond task-specific networks. Segment Anything Model (SAM) \citep{kirillovSegmentAnything2023a}, pretrained on approximately one billion masks, has been fine-tuned for crack detection in UAV imagery \citep{egodawelaDeepLearningApproach2024}. CrackSAM \citep{geFinetuningVisionFoundation2024} and \citet{guoSegmentAnythingModelbased2025} further explored parameter-efficient adaptation using LoRA and adapters. \citet{rostami2026segment} updated only 41K normalization parameters (0.046\% of SAM-ViT-Base) in Segment Any Crack, achieving strong zero-shot generalization while outperforming full fine-tuning. SAM2 \citep{raviSAM2Segment2024} further replaces SAM's ViT encoder with the hierarchical Hiera backbone \citep{ryaliHieraHierarchicalVision2023}, providing multiscale features and improved efficiency; however, its use for crack analysis remains limited \citep{chenSECrackSegHighAccuracyCrack2025}. Overall, these studies mainly focus on predicting accurate crack masks rather than condition-assessment quantities.

Beyond segmentation, studies have combined crack extraction with geometric measurement. Early methods relied on rule-based binarization and image processing to estimate crack length and width
\citep{fujitaRobustAutomaticCrack2011, adhikariImagebasedRetrievalConcrete2014, kimConcreteCrackIdentification2017}, however, hand-crafted crack extraction is sensitive to illumination, surface texture, and clutter. With advances in deep learning, learned segmentation increasingly replaced these procedures \citep{leeEstimationCrackWidth2020}. \citet{yangAutomaticPixelLevelCrack2018} used an FCN followed by medial-axis analysis to estimate crack length and width, although measurements remained in pixel units and were limited by whole-pixel resolution. \citet{niZernikemomentMeasurementThincrack2019} obtained subpixel widths using a dual-scale CNN and Zernike moments, while required high computational cost and was validated only under laboratory conditions. \citet{kangHybridPixellevelConcrete2020} combined Faster R-CNN, level-set segmentation, and distance-transform measurement, introducing error propagation across multiple stages. More recently, \citet{dingCrackDetectionQuantification2023} integrated transformer segmentation, edge-skeleton width estimation, sub-0.2~mm boundary refinement, and UAV-based 3D localization, whereas relied on fixed camera-specific pre-calibration. Thus, reliable conversion from segmentation masks to physical measurements under varying imaging conditions remains challenging.

Fewer studies have extended automated crack analysis to severity assessment. \citet{wangAutomatedCrackSeverity2021} classified ballastless track-slab crack severity directly from images. However, the study does not provide physical crack width or explicit inspection thresholds and required retraining when severity definitions changed. \citet{haAssessingSeverityRoad2022} classified five road-crack types and severity levels using SqueezeNet, dual U-Nets, and MobileNet-SSD, achieving 91.2\% accuracy, while relying on geometric proxies rather than calibrated physical widths. Consequently, existing severity methods are formulated as discrete classification problems, making their predictions difficult to trace to measurable crack characteristics and regulatory thresholds.

In summary, prior work has progressively advanced from crack segmentation to geometric quantification and severity prediction, however, these components remain disconnected. Key gaps include: (i) limited evaluation of segmentation robustness and zero-shot generalization across diverse crack scenes, (ii) geometric measurements that remain pixel-based or require fixed imaging calibration, and (iii) severity predictions that are not directly traceable to physical width and spatial-extent criteria used in inspection practice. This study addresses these gaps by: (i) systematically evaluating the accuracy, efficiency, and zero-shot generalization of adapted SAM2 variants, including the effects of prompting and image resolution; (ii) converting segmentation masks into validated physical crack-width and spatial-extent measurements using distance transforms and ground sampling distance (GSD), while establishing a resolution criterion for reliable width estimation; and (iii) integrating maximum crack width and spatial extent into a continuous, standards-informed Crack Severity Index (CSI).

\section{Methodology} \label{sec:methodology}

\begin{figure}[h]
    \centering
    \includegraphics[width=0.95\textwidth]{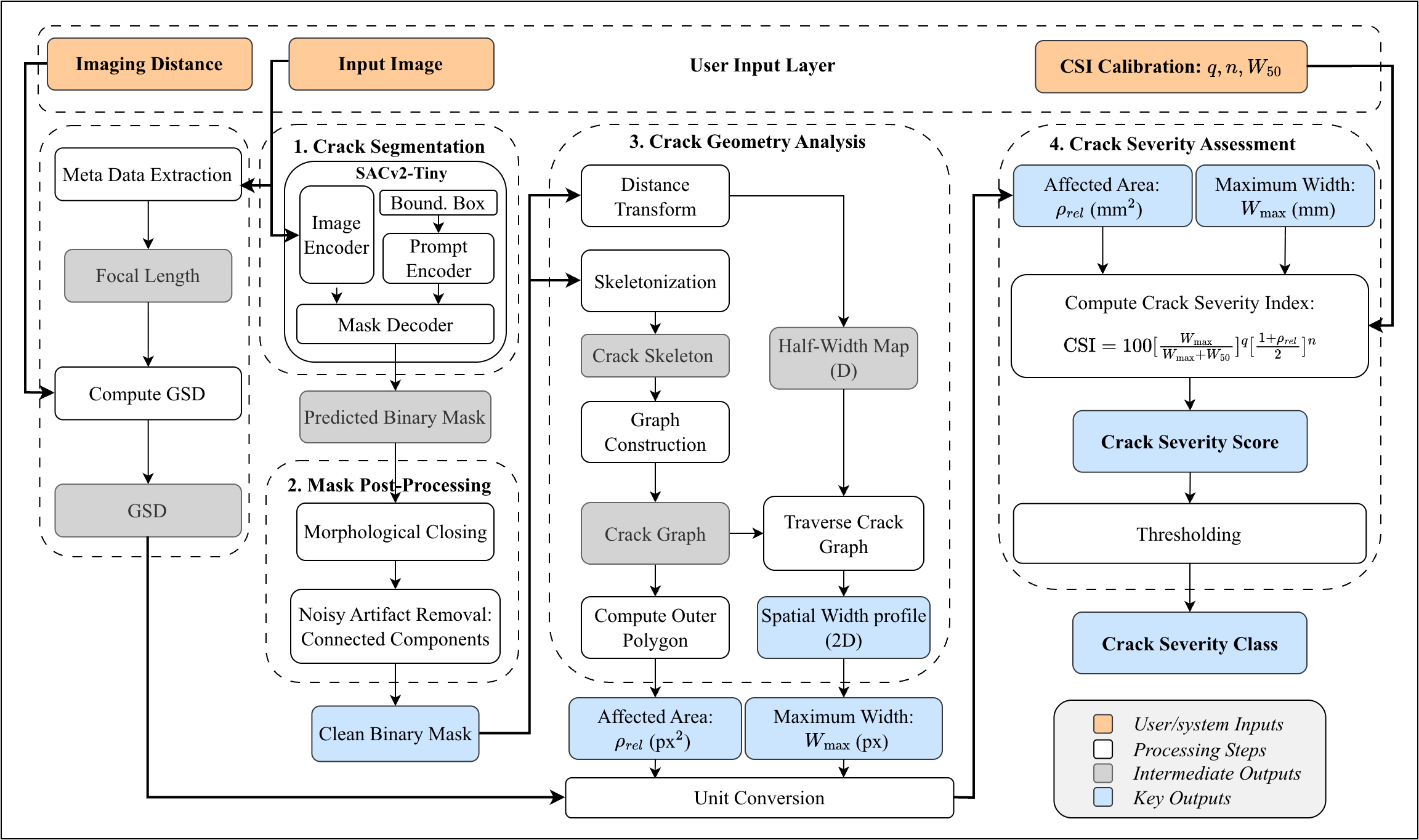}
    \caption{Overview of the proposed end-to-end crack assessment framework, Inspect Any Crack (IAC). Image metadata
             and imaging distance provide GSD calibration, while segmentation and mask
             post-processing produce a clean binary mask. Geometric analysis extracts maximum
             crack width and affected area, which are converted to physical units and combined
             by the calibrated Crack Severity Index (CSI) to determine the severity score and class.}
    \label{fig:pipeline}
\end{figure}

The proposed image-based framework integrates crack segmentation, geometric measurement,
physical calibration, and quantitative severity assessment. An adapted vision foundation model
first predicts a binary crack mask, which is post-processed and skeletonized to compute local
widths and area measures in pixels. These measurements are converted to physical units using a
reference object or physics-based ground sampling distance (GSD) and combined into a continuous
crack severity index. Figure~\ref{fig:pipeline} summarizes the end-to-end framework.
The following sections describe its six components: crack segmentation, geometric measurement, unit
conversion, severity assessment, datasets, and evaluation metrics.

\subsection{Crack Segmentation}

For robust and efficient crack segmentation, this study leverages Segment Anything Model~2 (SAM2), a state-of-the-art vision foundation model for segmentation \citep{raviSAM2Segment2024}, using the normalization-only fine-tuning method introduced by \citet{rostami2026segment}. SAM2 replaces SAM's monolithic vision transformer with the hierarchical Hiera encoder \citep{ryaliHieraHierarchicalVision2023}, providing multiscale features with improved computational efficiency. Its Tiny, Small, Base Plus, and Large variants offer different accuracy-efficiency trade-offs, therefore, all four variants were evaluated to identify the backbone providing the best balance of crack segmentation accuracy and inference speed.
The fine-tuning method updates only the affine scale and shift parameters of LayerNorm modules in the Hiera encoder and mask decoder, while all remaining parameters are frozen. This reduces trainable parameters, training cost, and overfitting risk while obtaining strong adaptation
\citep{rostami2026segment}. The approach is also consistent with \citet{zhaoTuningLayerNormAttention2023a}, which demonstrated that adapting channel-wise feature statistics can reduce domain discrepancy.
For training, images and masks were resized to $256\times256$ pixels, and the composite binary-cross-entropy and Dice loss of \citet{rostami2026segment} was employed. Optimization used AdamW with a learning rate of $5\times10^{-4}$, weight decay of $5\times10^{-5}$, cosine learning-rate scheduling, and a batch size of 2. For an $H\times W$ input image, the segmentation module predicts a binary mask
$\boldsymbol{M}\in{0,1}^{H\times W}$, where foreground pixels represent the detected crack regions. All segmentation components were implemented in PyTorch, and experiments were conducted on NVIDIA RTX~6000 Ada Generation GPUs.

\subsection{Crack Geometric Measurement}

\begin{figure}[h]
    \centering
    \includegraphics[width=0.8\textwidth]{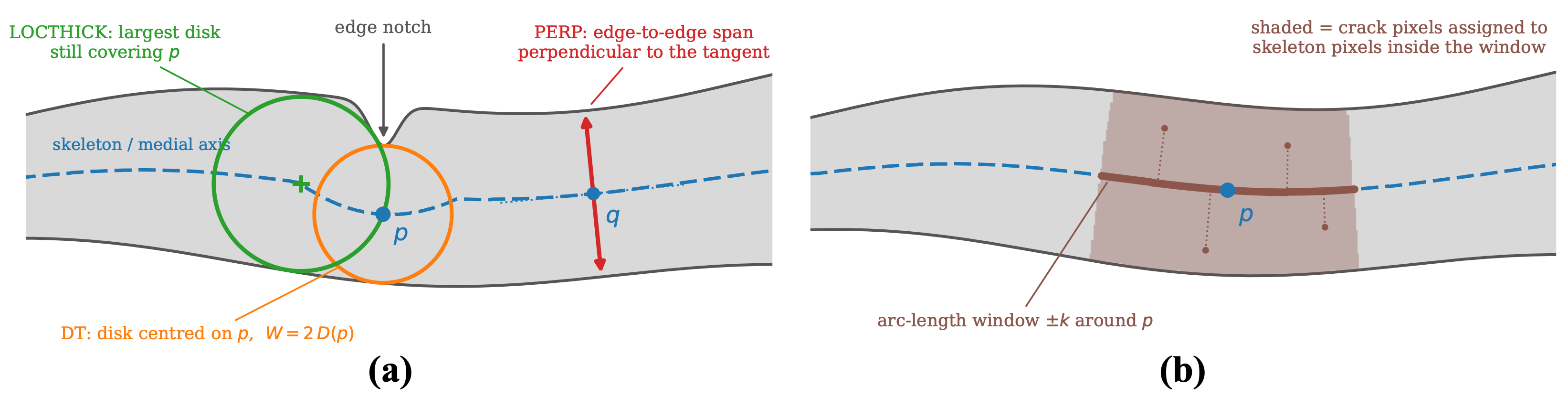}
    \caption{Schematic comparison of four local crack-width estimators at skeleton pixel
             $\boldsymbol{p}$: (a) distance transform (DT) in orange, local thickness (LocThick) in green, and
             perpendicular edge-pair scanning (PERP) in red, and (b) local windowed pixel counting (LocalPix) in brown.}
    \label{fig:width_estimator_schematic}
\end{figure}

The predicted mask is converted into geometric descriptors through post-processing, skeletonization, local-width estimation, and spatial-extent analysis. Morphological closing bridges small discontinuities, followed by removal of small few-pixel connected components as noise. The crack centerline is then extracted using the Lee thinning \citep{leeBuildingSkeletonModels1994} algorithm.
Due to the irregular and branching geometries of cracks,  four width estimators were evaluated under identical segmentation and calibration conditions:
\textbf{(1) Distance transform (DT)} \citep{meijsterGeneralAlgorithmComputing2000}: for a skeleton pixel $\boldsymbol{p}$, the Euclidean distance to the background, $D(\boldsymbol{p})$, represents the local half-width, giving $W(\boldsymbol{p})=2D(\boldsymbol{p})$.
\textbf{(2) Perpendicular edge-pair scanning (PERP)} \citep{liSemiautomaticCrackWidth2024}: local skeleton orientation is estimated by principal component analysis within an arc-length window, and rays projected along the normal intersect the interpolated mask boundary, where the edge-to-edge distance defines the local width.
\textbf{(3) Local thickness (LocThick)} \citep{hildebrand1997new}: width is defined as the diameter of the largest disk contained within the crack and containing $\boldsymbol{p}$. As
the disk need not be centered at $\boldsymbol{p}$, the estimate may reflect a nearby wider region.
\textbf{(4) Local windowed pixel counting (LocalPix)}: a localized variant of the area-over-length approach in \citet{yangAutomaticPixelLevelCrack2018}, where crack area within an adaptive arc-length window is divided by the corresponding skeleton length.
Figure~\ref{fig:width_estimator_schematic} illustrates the four methods, which were compared against caliper-referenced widths to select the lowest-error estimator. 

Furthermore, an 8-connected graph is constructed from the skeleton, with endpoints and junctions represented as nodes and connecting chains as edges. Graph traversal provides a spatial width profile, from which the maximum crack width, $W_{\max}$, is extracted.
Two area descriptors are also computed. The segmented crack area, $A_{\mathrm{crack}}$, is the foreground-pixel count, while the affected area, $A$, is the convex hull of the outermost skeleton endpoints and represents the surface footprint occupied by the crack network (Figure~\ref{fig:csi_components}). Their normalized forms are
$A_{\mathrm{rel}}=A_{\mathrm{crack}}/(HW)$ and $\rho_{\mathrm{rel}}=A/(HW)$.
Together, $W_{\max}$, $A_{\mathrm{rel}}$, and $\rho_{\mathrm{rel}}$ characterize both local crack severity and spatial extent. All geometric analyses were implemented in Python using SciPy and scikit-image.

\subsection{Unit Conversion through GSD Calibration} \label{subsec:unit_conversion}

Geometric measurements obtained from the mask are expressed in pixels and must be
converted to physical units using the image resolution
$r$ (mm/px) to support engineering interpretation and the downstream
concrete crack severity assessment:
 $W_{\max}^{\mathrm{mm}}=rW_{\max}^{\mathrm{px}}$ and
$A^{\mathrm{mm}^{2}}=r^{2}A^{\mathrm{px}^{2}}$, respectively.
To derive $r$, two calibration methods are used. If an object of known length is visible in the image,
$r$ is its physical length divided by its measured pixel length.
Without a reference object, $r$ is derived through Ground Sampling Distance (GSD) from the pinhole-camera model using stand-off distance
and image metadata \citep{jahanshahi2013new, lee2026exploratory}. Under this model
\citep{LuhmannRobsonKyleBoehm+2020}:
\vspace{-16pt}
\begin{equation} \label{eq:gsd}
GSD = \frac{p_s \cdot d}{f} \;,
\end{equation}
where $f$ is the actual focal length (mm) and $p_s$ is the sensor pixel pitch (mm/px).
The 35-mm-equivalent focal length avoids dependence on actual focal length and pixel pitch,
which vary among smartphone camera modules. For a sensor with aspect ratio $AR$
\citep{jahanshahi2013new, lee2026exploratory, cho2018effects}:
\vspace{-28pt}
\begin{equation} \label{eq:gsd_iphone}
r = GSD = \frac{43.27\,\, d}
         {f_{35}N_{\mathrm{short}}\sqrt{AR^{2}+1}} \;,
\end{equation}
where $d$ is the perpendicular stand-off distance (mm), $f_{35}$ is the 35-mm-equivalent
focal length from the image EXIF metadata, and $N_{\mathrm{short}}$ is the pixel count along the
shorter sensor dimension. For the 4:3 sensors of the iPhone~16 Pro Max,
$r\approx25.96d/(f_{35}N_{\mathrm{short}})$. The image-specific $f_{35}$ accommodates the
ultra-wide, main, and telephoto modules. 
When both calibrations are available, a discrepancy exceeding 10\% triggers review of stand-off
distance and camera alignment or the reference object.

\subsection{Crack Severity Assessment} \label{subsec:crack_severity}

\begin{figure}[h]
    \centering
    \includegraphics[width=0.65\textwidth]{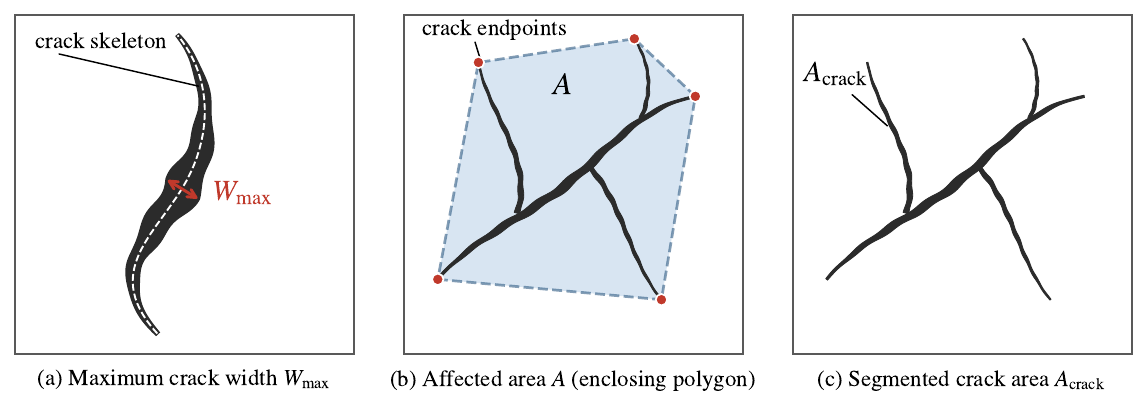}
             \caption{Geometric components of the Crack Severity Index (Eq.~\ref{eq:csi_full}):
             (a) maximum width $W_{\max}$, the largest local width normal to the crack
             skeleton, (b) affected area $A$, the polygon enclosing the outermost skeleton
             endpoints and representing network extent, and (c) segmented crack area
             $A_{\mathrm{crack}}$, the foreground area of the predicted mask. The normalized
             measures are $\rho_{rel}=A/(HW)$ and $A_{rel}=A_{\mathrm{crack}}/(HW)$.}
    
    \label{fig:csi_components}
\end{figure}

Concrete inspection regulations characterize surface-crack severity by maximum opening
width. However, width alone does not capture the additional risk posed by branching cracks,
including pathways for water ingress. This study therefore proposes a continuous Crack Severity
Index (CSI) that combines maximum opening width and spatial extent. Specifically, it integrates
maximum crack width, affected-area ratio, and segmented crack-area ratio, as depicted in Figure~\ref{fig:csi_components}:
\begin{equation} \label{eq:csi_full}
\mathrm{CSI}=100
\left(\frac{W_{\max}}{W_{\max}+W_{50}}\right)^{q}
\left(\frac{1+\rho_{rel}}{2}\right)^{n}
\left(\frac{1+A_{rel}}{2}\right)^{m} \;.
\end{equation}
Where $W_{\max}$ is expressed in mm, and $W_{50}$ is one-half of the critical crack width specified by the applicable inspection standard. The affected-area ratio
$\rho_{rel}=A/(HW)$ is the affected area normalized by the image area, while
$A_{rel}=A_{\mathrm{crack}}/(HW)$ is the normalized segmented crack area, with $H$ and $W$
denoting the image height and width. Positive exponents $q$, $n$, and $m$ control sensitivity
to crack opening, affected area, and segmented occupancy, respectively.

The CSI satisfies four requirements: (i) it is bounded within $[0, 100]$ and interpretable, (ii) it is
non-decreasing with crack width and spatial extent, (iii) it equals zero at zero width but
remains positive for a wide crack with a small footprint, and (iv) its width response is
anchored to an engineering reference. It is therefore expressed as
$100f_W(W_{\max})f_\rho(\rho_{rel})f_A(A_{rel})$, with each factor dimensionless
and bounded by unity.
The width factor, $f_W(W)=W/(W+W_{50})$, maps nonnegative widths to $[0,1)$ and
provides the half-response $f_W(W_{50})=1/2$. 
This saturating form retains sensitivity to small openings while bounding large widths.
Direct multiplication by $\rho_{rel}$ or $A_{rel}$ would nullify the CSI when either ratio
is zero. Both are therefore mapped as $g(x)=(1+x)/2$, increasing from $g(0)=1/2$
to $g(1)=1$. The extent measures can thus modify the width-based assessment without
eliminating its contribution.
These mappings ensure boundedness and monotonicity. For $W_{\max}\geq0$ and $q>0$,
$0\leq f_W^q<1$ and $\mathrm{d}f_W/\mathrm{d}W=W_{50}/(W+W_{50})^2>0$.
Likewise, $2^{-n}\leq g(\rho_{rel})^n\leq1$ and $2^{-m}\leq g(A_{rel})^m\leq1$ for $n,m>0$,
and both extent factors have nonnegative derivatives. Consequently, CSI equals zero
only when $W_{\max}=0$, increases with all three inputs, and approaches 100 only as crack
width becomes unbounded and both extent measures approach unity.
Thus, $0\leq\mathrm{CSI}\leq100$ for all positive $q$, $n$, and $m$.
The exponents $q$, $n$, and $m$ control the relative influence of crack opening, affected
extent, and segmented occupancy. Sensitivity analyses were conducted to assess their effects. 

\subsection{Datasets}

\begin{figure}[h]
    \centering
    \includegraphics[width=\textwidth]{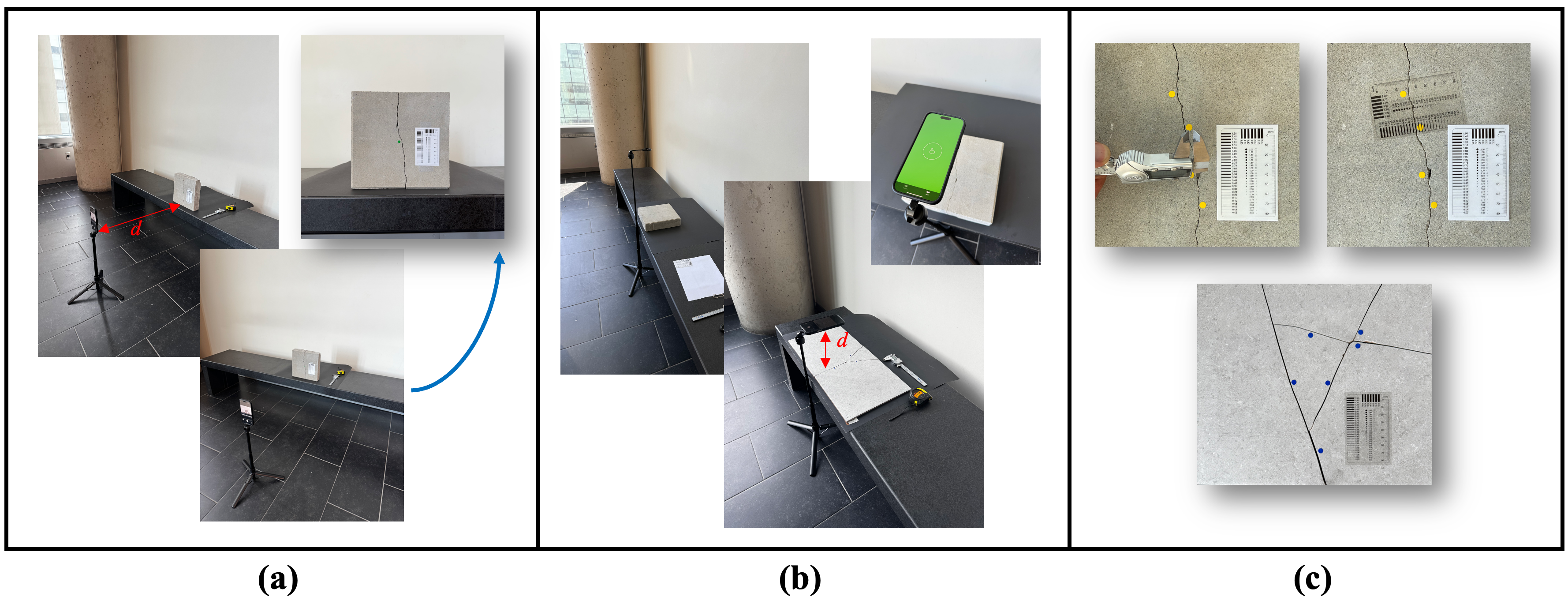}
             \caption{Image acquisition and reference-width measurement for the concrete crack dataset collected for this study: (a) iPhone 16 Pro Max setup and representative image for the variable imaging-distance (and GSD) experiment, (b) fixed-distance ($d=20$cm) imaging of linear and web cracks, and (c) representative digital-caliper and crack-width-comparator measurements. Yellow and blue markers indicate caliper-referenced locations on the linear and web cracks, respectively.}
    \label{fig:sample_acquisition}
\end{figure}

OmniCrack30k was used for segmentation training, validation, and in-distribution testing
\citep{benzOmniCrack30kBenchmarkCrack2024a}. It aggregates more than 20 crack datasets and
approximately 30,000 image-mask pairs from concrete, asphalt, ceramic, masonry, and steel
surfaces, with diverse resolutions, fields of view, illumination, textures, and crack morphologies.
Zero-shot generalization was evaluated on three datasets excluded from training: Road420,
Facade390 \citep{geFinetuningVisionFoundation2024}, and Concrete3k \citep{liRealtimeHighresolutionNeural2023,wangAutomaticConcreteCrack2022}.
For geometric measurement validation, a separate dataset was collected for this study, comprising 21 images of 11 concrete crack samples acquired using an iPhone~16 Pro Max (Fig.~\ref{fig:sample_acquisition}).
One sample with a caliper-verified width of $w^*=1.5$\,mm was imaged at eleven stand-off
distances to evaluate ground sampling distance (GSD) effects on segmentation continuity and width error
(Fig.~\ref{fig:sample_acquisition}(a), Table~\ref{table:width_methods_s15}).
The remaining ten samples comprised seven linear and three web (branching) cracks, providing 41
caliper-verified widths from 0.5 to 7.6\,mm (Fig.~\ref{fig:sample_acquisition}(b,c)).
The ten samples were imaged at $d=20$\,cm for the end-to-end validation described in
Section~\ref{subsec:measurement_results} and in
Table~\ref{tab:width_comparison_all}.
Table~\ref{table:datasets} summarizes the datasets and their roles.

\begin{table}[h]
\small
\renewcommand{\arraystretch}{1.15}
\centering
\footnotesize
\caption{Datasets used for segmentation development, zero-shot evaluation, and geometric measurement validation.}
\label{table:datasets}
\begin{tabular}{l c c c c c}
    \hline\hline
    Dataset          & Purpose              & Train  & Val    & Test  & \multicolumn{1}{c}{Image Size}\\
    \hline
    OmniCrack30k     & Segmentation         & 22,158 & 13,277 & 4,582 & $[81, 4608] \times [116, 4608]$ \\
    Road420          & Zero-shot            & ---    & ---    & 420  & $448\times 448$ \\
    Facade390        & Zero-shot            & ---    & ---    & 390   & $448\times 448$ \\
    Concrete3k       & Zero-shot            & ---    & ---    & 3,000 & $500\times 500$ \\
    \multirow{2}{*}{Collected dataset (this study)} &
    \multirow{2}{*}{Measurement} &
    \multirow{2}{*}{---} &
    \multirow{2}{*}{---} &
    \multirow{2}{*}{21} &
    $3024\times3024$, $4284\times4284$ \\
    & & & & & $5712\times4284$ \\
    \hline\hline
\end{tabular}
\end{table}

\subsection{Evaluation Metrics}

The framework was evaluated in terms of segmentation accuracy, computational efficiency, and
geometric measurement accuracy.
\textit{Segmentation accuracy} was evaluated using pixel-level Precision, Recall, F1-score, and
Intersection over Union (IoU):
\vspace{-9pt}
\begin{equation}
\mathrm{Precision} = \frac{TP}{TP+FP} \;, \qquad
\mathrm{Recall} = \frac{TP}{TP+FN} \;,
\end{equation}
\begin{equation}
\mathrm{F1} =
\frac{2\,\mathrm{Precision}\,\mathrm{Recall}}
     {\mathrm{Precision}+\mathrm{Recall}} \;, \qquad
\mathrm{IoU} = \frac{TP}{TP+FP+FN} \;,
\end{equation}
Where $TP$, $FP$, and $FN$ denote correctly detected crack pixels, false-positive background pixels,
and missed crack pixels, respectively. F1-score and IoU were the main metrics for in-distribution
evaluation on OmniCrack30k and zero-shot evaluation on Road420, Facade390, and Concrete3k.
\textit{Computational efficiency} was measured using total parameters, and inference throughput in frames per second (FPS), reflecting field and edge deployment
requirements. To jointly evaluate segmentation accuracy and computational efficiency, this study defines the composite performance score $S_{\mathrm{perf}}$ as the geometric mean of the normalized F1-score and FPS:
\vspace{-9pt}
\begin{equation}
    S_{\mathrm{perf}} = \sqrt{\,\frac{\text{F1}}{\text{F1}_{\max}} \times \frac{\text{FPS}}{\text{FPS}_{\max}}},
    \label{eq:performance_score}
\end{equation}
where $\text{F1}_{\max}$ and $\text{FPS}_{\max}$ are the best observed values. Thus,
$S_{\mathrm{perf}}\in(0,1]$, with 1 indicating the highest accuracy and throughput. The geometric
mean penalizes imbalance between the two criteria.
\textit{Geometric measurement accuracy} is evaluated against caliper-verified widths at discrete
crack locations. At location $i$, $w_i^{*}$ and $\hat{w}_i$ denote the reference and estimated
widths in mm. Absolute Error (AE) and Relative Error (RE) are:
\vspace{-20pt}
\begin{equation}
\mathrm{AE}_i = \left|\hat{w}_i-w_i^{*}\right| \;, \qquad
\mathrm{RE}_i =
\frac{\left|\hat{w}_i-w_i^{*}\right|}{w_i^{*}}\times100\% \;.
\end{equation}
\vspace{-3pt}
For $N$ reference locations, Mean Absolute Error (MAE) and Mean Relative Error (MRE) are:
\vspace{-10pt}
\begin{equation}
\mathrm{MAE} = \frac{1}{N}\sum_{i=1}^{N}\mathrm{AE}_i \;, \qquad
\mathrm{MRE} = \frac{1}{N}\sum_{i=1}^{N}\mathrm{RE}_i \;.
\end{equation}
MAE reports average physical error, while MRE enables comparison across crack widths.

\section{Results and Discussion} \label{sec:results}

\subsection{Input Prompt Ablation Studies} \label{subsubsec:input_ablation}

\begin{figure}[ht]
    \centering
    \includegraphics[width=0.65\textwidth]{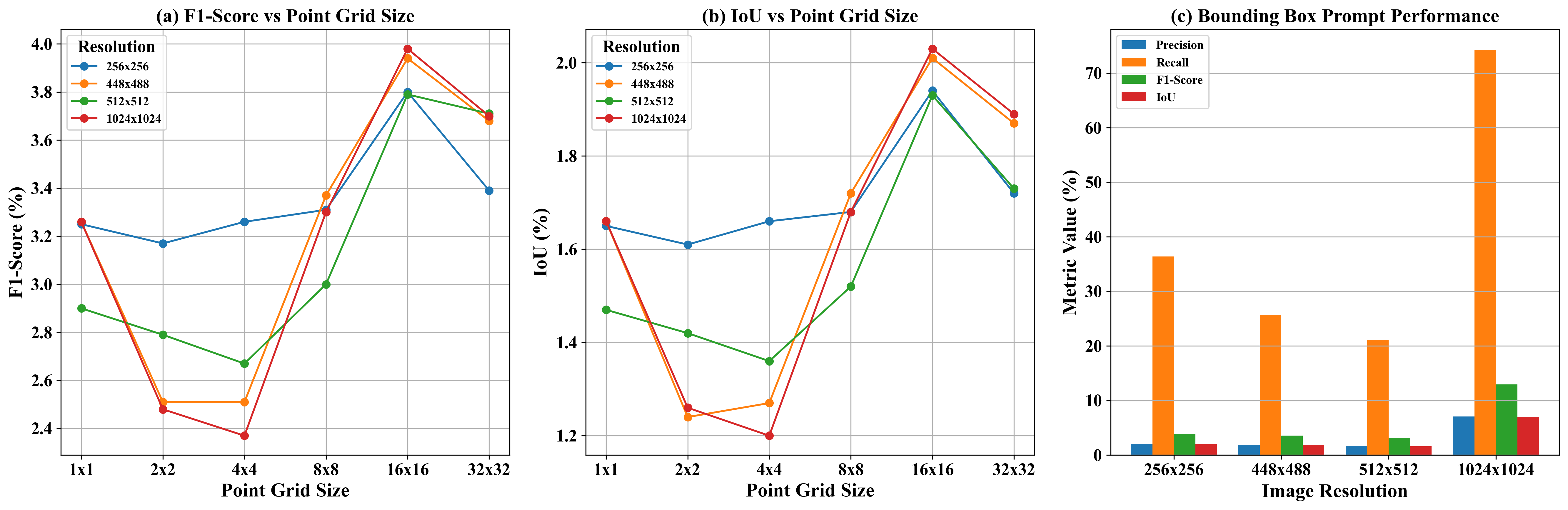}
    \includegraphics[width=0.3\textwidth]{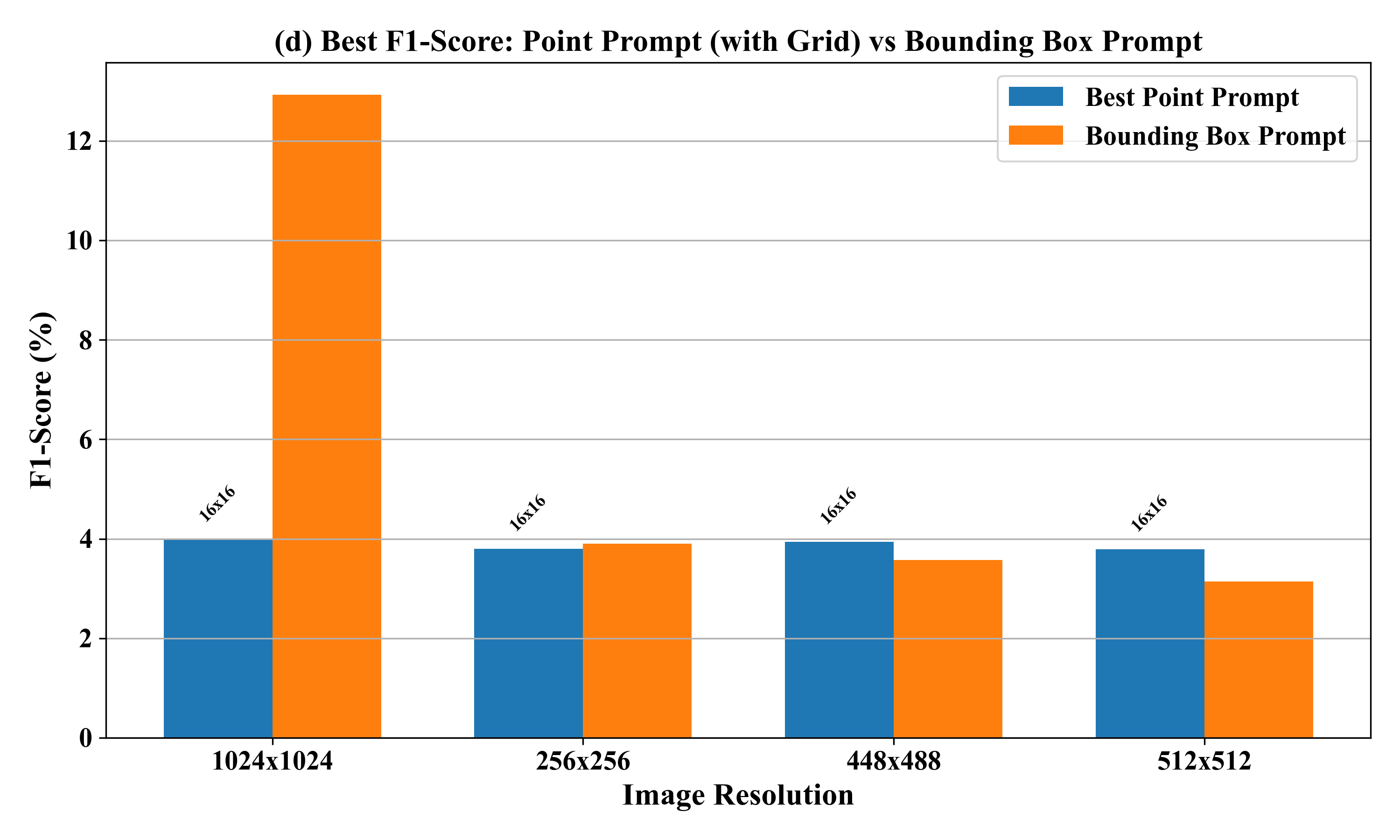}
    \caption{Effects of prompt type and input resolution on the segmentation performance
             of the untuned SAM2.1-Tiny model on Road420: (a) F1-score and (b) IoU for
             point grids of increasing density, (c) segmentation performance vs image resolution using bounding box prompt, and (d) comparison of the best point grid
             ($16\times16$) with a bounding box. Bounding-box prompts provide the highest
             F1-score and IoU at every evaluated resolution.}
    \label{fig:ablative_sam2_combined}
\end{figure}

\begin{figure}[ht]
    \centering
    \includegraphics[width=0.6\textwidth]{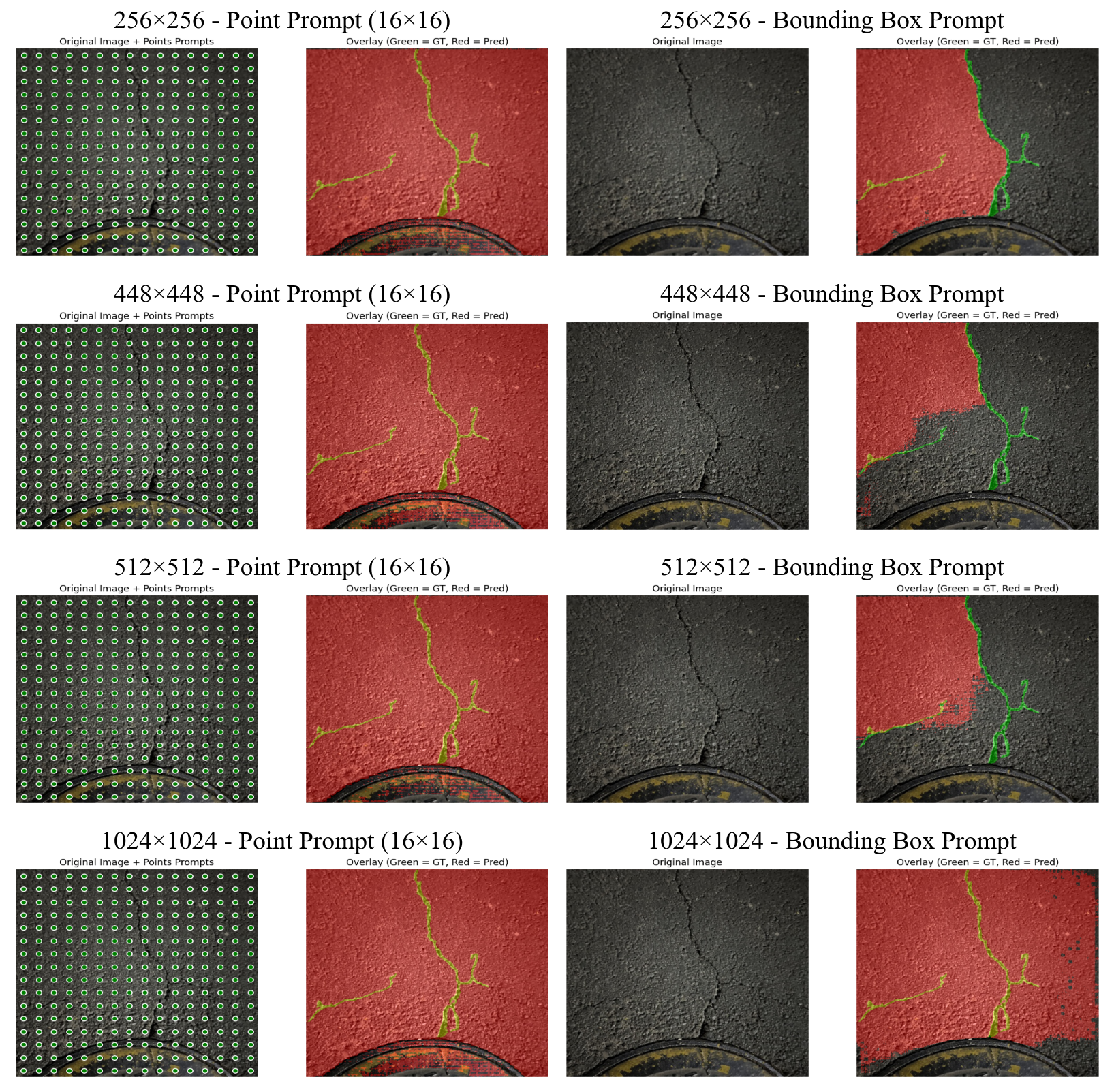}
    \caption{Qualitative comparison of predictions obtained with a $16\times16$ point
             grid and a bounding-box prompt at input resolutions from $256\times256$ to
             $1024\times1024$ pixels (top to bottom). Ground-truth and predicted crack
             regions are shown in green and red, respectively. Bounding-box prompts
             recover more continuous crack regions and produce fewer false positives at
             each resolution.}
    \label{fig:prompt_qualitative_comparison}
\end{figure}

Prior to fine-tuning, an ablation study evaluated prompt type and input resolution to the Segment Anything Model (SAM2) for crack
segmentation. The untuned SAM2-Tiny model was tested on Road420 using a bounding box around the input image or
point grids of increasing density at $256\times256$, $448\times448$, $512\times512$, and
$1024\times1024$ pixels. Figure~\ref{fig:ablative_sam2_combined} shows that bounding-box
prompts outperform all point-grid configurations at every resolution. Bounding-box performance
increases with resolution, reaching a 12.93\% F1-score and 6.91\% IoU at $1024\times1024$
pixels. At the same resolution, the best point-based prompt, a $16\times16$ grid, achieves only
3.98\% F1-score and 2.03\% IoU. Although denser point grids improve crack localization, a
substantial performance gap remains. Furthermore, the qualitative results in
Fig.~\ref{fig:prompt_qualitative_comparison} show that bounding boxes recover more
continuous crack regions with fewer false positives across all four resolutions.
Therefore, bounding boxes were used in all subsequent experiments. Although the untuned model
performs best at $1024\times1024$ pixels, increasing the linear resolution from 256 to 1024
raises the input pixel count 16-fold, increasing training memory and computation. The resolution
was therefore fixed at $256\times256$ pixels for fine-tuning. The low absolute scores by the untuned model also
show that effective prompts alone are insufficient for crack segmentation, motivating the
task-specific fine-tuning evaluated in the following subsection.

\subsection{Segmentation Performance} \label{subsec:seg_results}

To assess the effect of backbone capacity on segmentation accuracy and computational
efficiency, the four SAM2 variants, including Tiny, Small, Base Plus, and Large, were
fine-tuned on OmniCrack30k using the normalization-only fine-tuning method described previously.
All experiments used bounding-box prompts and $256\times256$-pixel inputs. 

\begin{figure}[ht]
    \centering
    \includegraphics[width=0.55\textwidth]{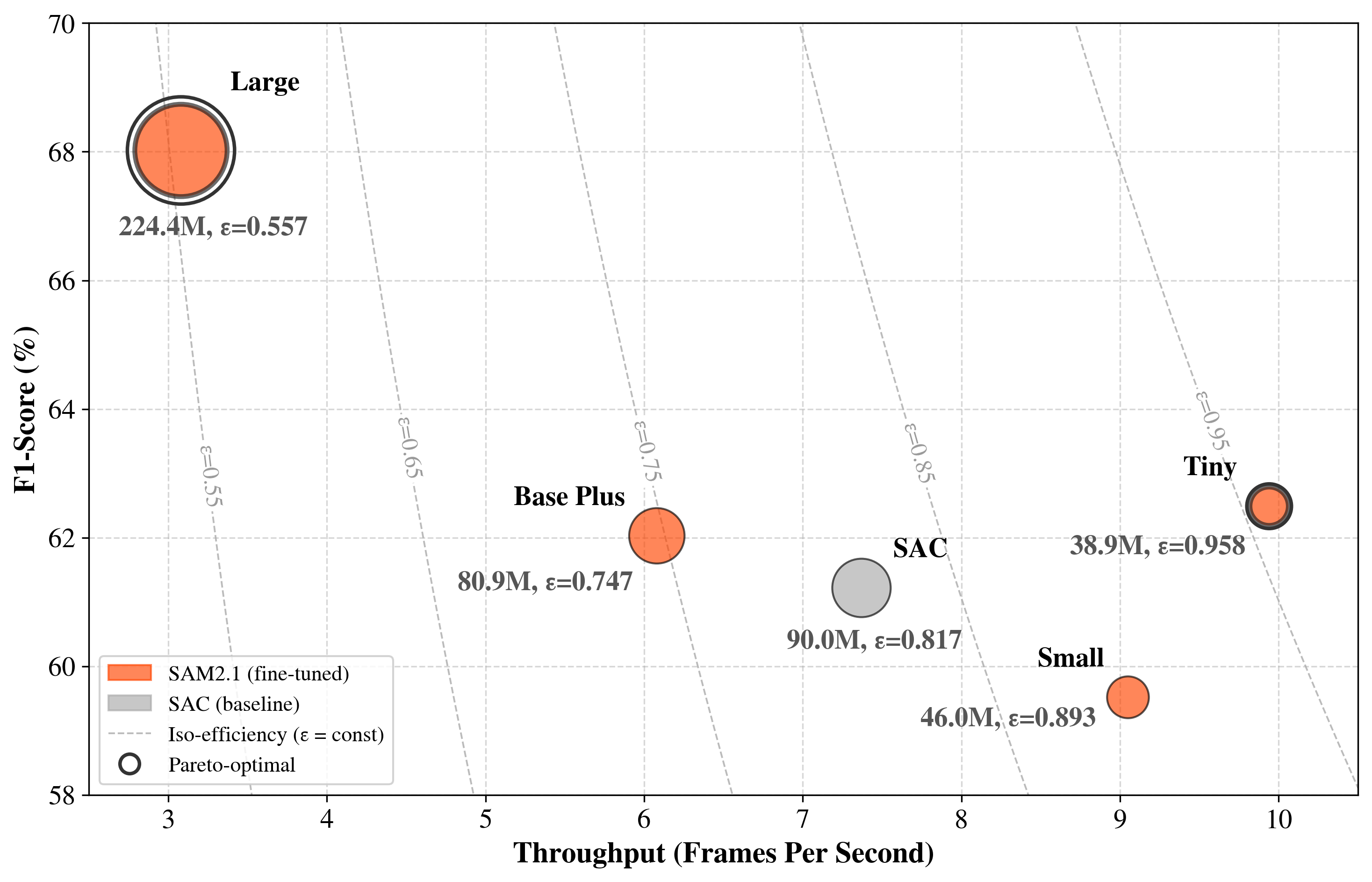}
     \caption{Accuracy-efficiency trade-off on OmniCrack30k, with bubble sizes proportional
             to the model size. Dashed contours denote constant performance scores
             $S_{\mathrm{perf}}$, and dark outer rings identify Pareto-optimal models.
             SACv2-Tiny exceeds SAC in F1-score at 9.94 frames/s with 56.8\% fewer parameters.
             Large attains the highest F1-score while providing 69.0\% lower throughput.}
    \label{fig:sam2_f1_vs_fps}
\end{figure}

Table~\ref{tab:sam2_results_combined} and Fig.~\ref{fig:sam2_f1_vs_fps} show that segmentation
accuracy does not increase monotonically with model size under normalization-only fine-tuning. Large
achieves the highest accuracy (68.02\% F1-score and 51.54\% IoU), whereas Tiny reaches
62.49\% F1-score and 45.45\% IoU with 38.9 million parameters, updating only 29.8 thousand
normalization parameters. Base Plus performs similarly to Tiny (62.03\% F1-score and
44.96\% IoU), while Small achieves 59.52\% F1-score and 42.37\% IoU despite having more
parameters. Thus, adaptation performance depends not only on parameter count but also on
the distribution of normalization parameters and the pretrained representations
\citep{tayScaleEfficientlyInsights2021, linSelectingLargeLanguage2024}. Tiny also provides
the highest throughput on an NVIDIA RTX 3070 GPU-equipped machine with 9.94 FPS, followed
by Small (9.05 FPS), SAC \citep{rostami2026segment} (7.37 FPS), Base Plus
(6.08 FPS), and Large (3.08 FPS). Compared with Tiny, Large improves F1-score by
5.53 percentage points but reduces throughput by 69.0\% and uses 5.77 times as many total
parameters. This trade-off is relevant to inspection systems constrained by onboard memory, and processing capacity.
The dimensionless performance score in Eq.~\eqref{eq:performance_score} balances segmentation
accuracy and inference speed. Table~\ref{tab:sam2_results_combined} shows that Tiny achieves
the highest score ($S_{\mathrm{perf}}=0.959$), indicating that its higher throughput outweighs
its 5.53-percentage-point F1-score deficit relative to Large. Combined with its lower parameter
count, this result supports selecting SAM2-Tiny as the backbone of the proposed model, hereafter
the Segment Any Crackv2-Tiny (SACv2-Tiny). On OmniCrack30k, SACv2-Tiny improves upon the SAC baseline
by 1.27 percentage points in F1-score (62.49\% versus 61.22\%) and 1.32 percentage points in IoU
(45.45\% versus 44.13\%), while reducing the total parameter count from 90.0 million to
38.9 million (56.8\%) and increasing throughput from 7.37 to 9.94 FPS (34.9\%) on the
same hardware. Thus, SACv2-Tiny improves both segmentation accuracy and computational efficiency
relative to SAC.

The zero-shot results in Table~\ref{tab:sam2_results_combined} assess generalization beyond
the training domain. Across Road420, Facade390, and Concrete3k, SACv2-Tiny achieves mean
zero-shot F1-score and IoU values of $69.89_{\pm{3.24}}$\% and
$53.82_{\pm{3.89}}$\%, respectively, compared with $67.20_{\pm{6.05}}$\% and
$50.93_{\pm{7.07}}$\% for SAC. Although Large achieves the highest mean accuracy
($70.96_{\pm{1.71}}$\% F1-score and $55.02_{\pm{2.04}}$\% IoU), SACv2-Tiny yields the
highest zero-shot performance score ($S_{\mathrm{perf}}^{\mathrm{zs}}=0.992$) and
lower cross-dataset variation. The results support SACv2-Tiny as the segmentation backbone
for crack measurement and severity assessment, balancing segmentation quality and
computational cost.

\begin{table*}[t]
\centering
\caption{Accuracy, efficiency, and generalization of the fine-tuned
         SAM2 variants and SAC baseline: (a) OmniCrack30k in-domain results and
         inference throughput, and (b) zero-shot performance on Road420, Facade390,
         and Concrete3k. Zero-shot averages are reported as mean $\pm$ standard
         deviation across the three datasets. $\dagger$ denotes Pareto-optimal model.}
\label{tab:sam2_results_combined}
\label{table:zeroshot}
\footnotesize
\renewcommand{\arraystretch}{1.2}

\textbf{(a) OmniCrack30k accuracy and computational efficiency}\par\smallskip
\begin{tabularx}{\textwidth}{l X X X X X X X X}
\hline\hline
\textbf{Model} & \textbf{Params (M)} & \textbf{Norm Params} &
\textbf{Prec.\ (\%)} & \textbf{Rec.\ (\%)} & \textbf{F1 (\%)} &
\textbf{IoU (\%)} & \textbf{FPS} & $S_{\mathrm{perf}}$ \\
\hline
SAC \citep{rostami2026segment} & 90.0 & 41K  & 64.16 & 58.60 & 61.22 & 44.13 & 7.37 & 0.817 \\
\hline
\textbf{Tiny$^\dagger$ (SACv2-Tiny)} & \textbf{38.9} & \textbf{29.8K} &
    \textbf{63.81} & \textbf{61.23} & \textbf{62.49} & \textbf{45.45} &
    \textbf{9.94} & \textbf{0.959} \\
Small               & 46.0  & 35.9K  & 61.37 & 57.78 & 59.52 & 42.37 & 9.05 & 0.893 \\
Base Plus           & 80.9  & 53.7K  & 69.58 & 55.97 & 62.03 & 44.96 & 6.08 & 0.747 \\
Large$^\dagger$     & 224.4 & 119.7K & 71.76 & 64.66 & 68.02 & 51.54 & 3.08 & 0.557 \\
\hline\hline
\end{tabularx}

\vspace{6pt}
\textbf{(b) Zero-shot generalization}\par\smallskip
\begin{tabularx}{\textwidth}{l *{9}{>{\centering\arraybackslash}X}}
    \hline\hline
    \multirow{2}{*}{Model} &
    \multicolumn{2}{c}{Road420} &
    \multicolumn{2}{c}{Facade390} &
    \multicolumn{2}{c}{Concrete3k} &
    \multicolumn{2}{c}{Avg. Zero-Shot} &
    \multirow{2}{*}{$S_{\mathrm{perf}}^{\mathrm{zs}}$} \\
    \cmidrule(lr){2-3}\cmidrule(lr){4-5}\cmidrule(lr){6-7}\cmidrule(lr){8-9}
    & F1 (\%) & IoU (\%) & F1 (\%) & IoU (\%) & F1 (\%) & IoU (\%) &
      F1 (\%) & IoU (\%) & \\
    \hline
    SAC & 64.22 & 47.30 & 61.74 & 44.68 & 75.63 & 60.82 &
        $67.20_{\pm{6.05}}$ & $50.93_{\pm{7.07}}$ & 0.838 \\
    \hline
    \textbf{Tiny}$^\dagger$ & 67.76 & 51.24 & 67.45 & 50.89 & 74.47 & 59.32 &
        $69.89_{\pm{3.24}}$ & $53.82_{\pm{3.89}}$ & \textbf{0.992} \\
    Small & 70.47 & 54.40 & 64.82 & 47.95 & 72.64 & 57.03 &
        $69.31_{\pm{3.30}}$ & $53.13_{\pm{3.82}}$ & 0.943 \\
    Base Plus & 62.44 & 45.40 & 75.02 & 60.02 & 71.13 & 55.20 &
        $69.53_{\pm{5.26}}$ & $53.54_{\pm{6.09}}$ & 0.774 \\
    Large & 72.53 & 56.90 & 68.58 & 52.18 & 71.77 & 55.97 &
        $70.96_{\pm{1.71}}$ & $55.02_{\pm{2.04}}$ & 0.557 \\
    \hline\hline
\end{tabularx}

\par\smallskip
\footnotesize
For panel (a), $S_{\mathrm{perf}}$ uses
$\mathrm{F1}_{\max}=68.02\%$ and $\mathrm{FPS}_{\max}=9.94$.
For panel (b), $S_{\mathrm{perf}}^{\mathrm{zs}}$ uses
$\mathrm{F1}_{\max}=70.96\%$ and the same $\mathrm{FPS}_{\max}$.
\end{table*}

\subsection{Width Estimator Selection and GSD Analysis} \label{subsec:gsd}

\begin{figure}
    \centering
    \begin{minipage}[c]{0.06\linewidth}\centering\rotatebox[origin=c]{90}{\small $d=10$\,cm}\end{minipage}
    \begin{minipage}[c]{0.65\linewidth}
        \includegraphics[width=\linewidth]{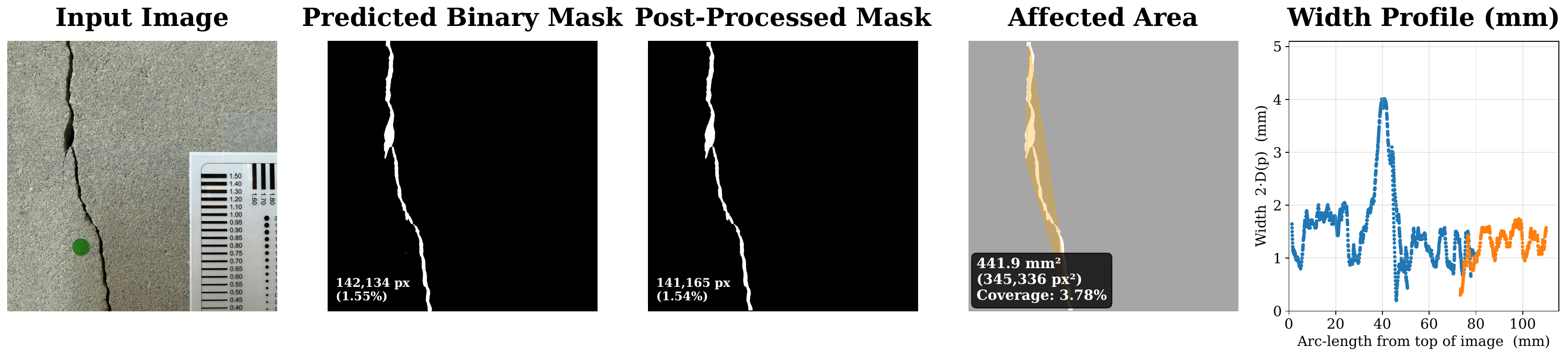}
    \end{minipage}\vspace{1pt}

    \begin{minipage}[c]{0.06\linewidth}\centering\rotatebox[origin=c]{90}{\small $d=30$\,cm}\end{minipage}
    \begin{minipage}[c]{0.65\linewidth}
        \includegraphics[width=\linewidth]{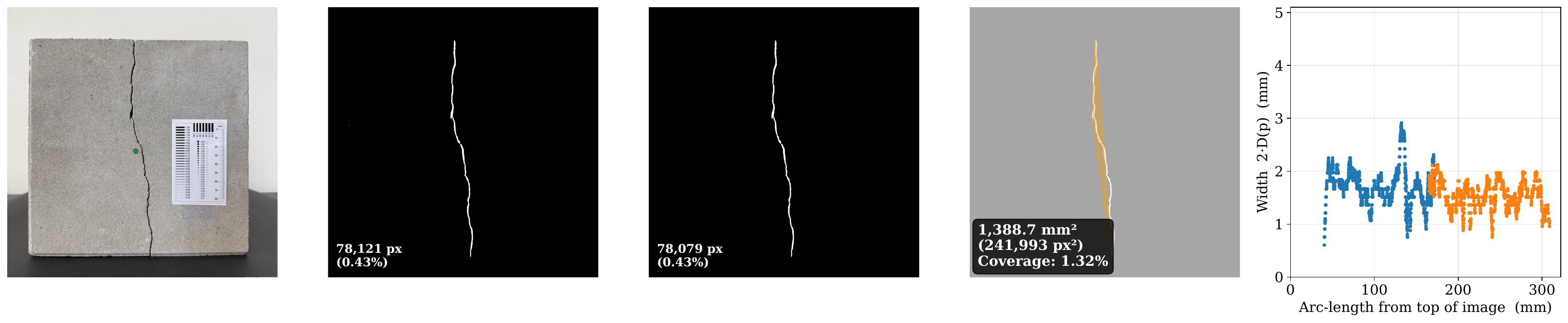}
    \end{minipage}\vspace{1pt}

    \begin{minipage}[c]{0.06\linewidth}\centering\rotatebox[origin=c]{90}{\small $d=60$\,cm}\end{minipage}
    \begin{minipage}[c]{0.65\linewidth}
        \includegraphics[width=\linewidth]{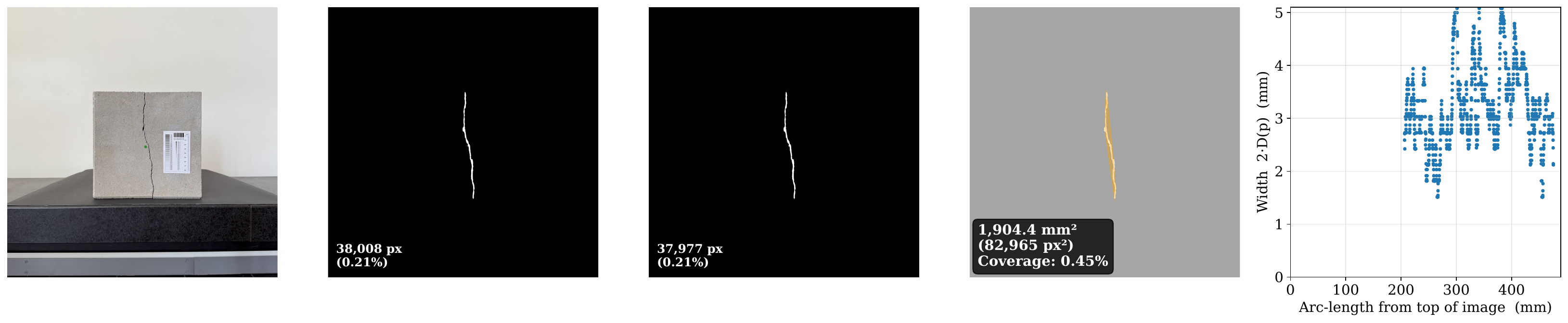}
    \end{minipage}\vspace{1pt}

    \begin{minipage}[c]{0.06\linewidth}\centering\rotatebox[origin=c]{90}{\small $d=120$\,cm}\end{minipage}
    \begin{minipage}[c]{0.65\linewidth}
        \includegraphics[width=\linewidth]{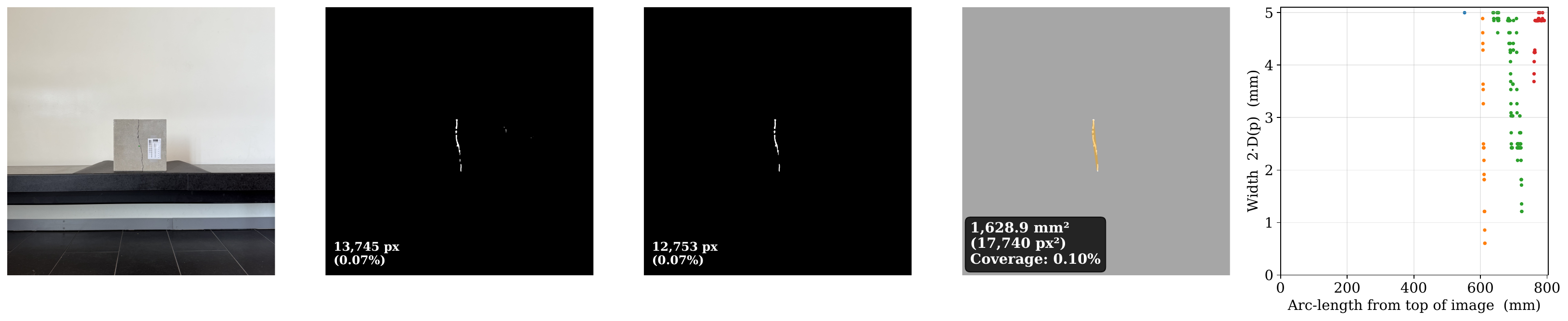}
    \end{minipage}
    \caption{Sample crack captures with reference width photographed at different distances. The crack width profile is obtained with distance transform. The green marker identifies the caliper reference
             ($w^*=1.5$\,mm). Increasing distance progressively fragments the mask and increases width error as crack visibility declines.}
    \label{fig:sample_15_distance_exp}
\end{figure}

Following segmentation, the binary crack mask is used to estimate physical crack width. This requires two considerations: (1) selecting an accurate mask-based width estimator and (2) determining a suitable ground sampling distance (GSD) for reliable measurement. For this purpose, a crack with a caliper-referenced width of $w^*=1.5,\mathrm{mm}$ was photographed at the same reference location from eleven stand-off distances ($d=10$-$120$,cm), spanning GSDs of $0.036$-$0.303$,mm/px (Figures~\ref{fig:sample_acquisition}-a and \ref{fig:sample_15_distance_exp}). The GSD of each image was computed from its EXIF metadata using Eq.~\eqref{eq:gsd_iphone}. Keeping the crack and measurement location fixed isolates image resolution as the main variable, enabling direct comparison of the width estimators and quantification of GSD-dependent measurement error.

\textbf{Width Estimation Algorithm.}
The four width estimators described previously were evaluated at the same reference location across all eleven images. As shown in Table~\ref{table:width_methods_s15}, DT achieved the lowest aggregate error (MAE $=1.88$,mm, mean RE $=125.5\%$), followed by LocalPix ($1.91$,mm, $127.3\%$), PERP ($2.13$,mm, $141.7\%$), and LocThick ($2.52$,mm, $167.9\%$). DT is comparatively less sensitive to boundary irregularities as it measures the shortest distance from the skeleton to the crack boundary. PERP can intersect oblique or irregular boundaries, LocThick may extend into neighboring wider regions, and LocalPix is sensitive to window size and local width variation, leading to greater overestimation. DT was therefore selected for the proposed framework, and all subsequent crack-width measurements use this estimator.

\begin{table}[h]
\centering
\caption{Comparison of four width estimators across eleven stand-off distances at a
fixed reference point ($w^*{=}1.5$\,mm). The table reports estimated width and relative
error by Distance Transform (DT), Perpendicular edge-pair scanning (PERP), Local thickness (LocThick), and Local windowed pixel counting (LocalPix). Bold values denote the best aggregate performance.}
\label{table:width_methods_s15}
\footnotesize
\renewcommand{\arraystretch}{1.1}
\setlength{\tabcolsep}{4pt}
\begin{tabular}{c c c c c c c c c c}
    \hline\hline
    \multirow{2}{*}{\shortstack{Dist.\\(cm)}} &
    \multirow{2}{*}{\shortstack{GSD\\(mm/px)}} &
    \multicolumn{2}{c}{DT} & \multicolumn{2}{c}{PERP} &
    \multicolumn{2}{c}{LocThick} & \multicolumn{2}{c}{LocalPix} \\
    \cmidrule(lr){3-4}\cmidrule(lr){5-6}\cmidrule(lr){7-8}\cmidrule(lr){9-10}
     & & $\hat{w}$ & RE & $\hat{w}$ & RE & $\hat{w}$ & RE & $\hat{w}$ & RE \\
     & & (mm) & (\%) & (mm) & (\%) & (mm) & (\%) & (mm) & (\%) \\
    \hline
    10  & 0.036 & 1.50 & 0.0   & 1.64 & 9.3   & 1.65 & 10.0  & 1.44 & 4.0   \\
    20  & 0.051 & 1.50 & 0.0   & 1.48 & 1.3   & 1.52 & 1.3   & 1.41 & 6.0   \\
    30  & 0.076 & 1.52 & 1.3   & 1.85 & 23.3  & 1.97 & 31.3  & 1.83 & 22.0  \\
    40  & 0.101 & 2.02 & 34.7  & 2.24 & 49.3  & 2.63 & 75.3  & 2.29 & 52.7  \\
    50  & 0.126 & 2.78 & 85.3  & 2.84 & 89.3  & 2.78 & 85.3  & 2.65 & 76.7  \\
    60  & 0.152 & 3.39 & 126.0 & 3.64 & 142.7 & 3.64 & 142.7 & 3.40 & 126.7 \\
    70  & 0.177 & 3.69 & 146.0 & 4.26 & 184.0 & 4.95 & 230.0 & 4.00 & 166.7 \\
    80  & 0.202 & 3.98 & 165.3 & 4.28 & 185.3 & 4.85 & 223.3 & 3.93 & 162.0 \\
    90  & 0.227 & 5.00 & 233.3 & 5.34 & 256.0 & 5.68 & 278.7 & 4.78 & 218.7 \\
    100 & 0.253 & 5.76 & 284.0 & 5.95 & 296.7 & 6.06 & 304.0 & 5.25 & 250.0 \\
    120 & 0.303 & 6.06 & 304.0 & 6.32 & 321.3 & 8.48 & 465.3 & 6.22 & 314.7 \\
    \hline
    \multicolumn{2}{l}{MAE (mm)} &
        \multicolumn{2}{c}{\textbf{1.88}} & \multicolumn{2}{c}{2.13} &
        \multicolumn{2}{c}{2.52} & \multicolumn{2}{c}{1.91} \\
    \multicolumn{2}{l}{Mean RE (\%)} &
        \multicolumn{2}{c}{\textbf{125.5}} & \multicolumn{2}{c}{141.7} &
        \multicolumn{2}{c}{167.9} & \multicolumn{2}{c}{127.3} \\
    \hline\hline
\end{tabular}
\end{table}

\begin{figure}
    \centering
    \includegraphics[width=0.7\linewidth]{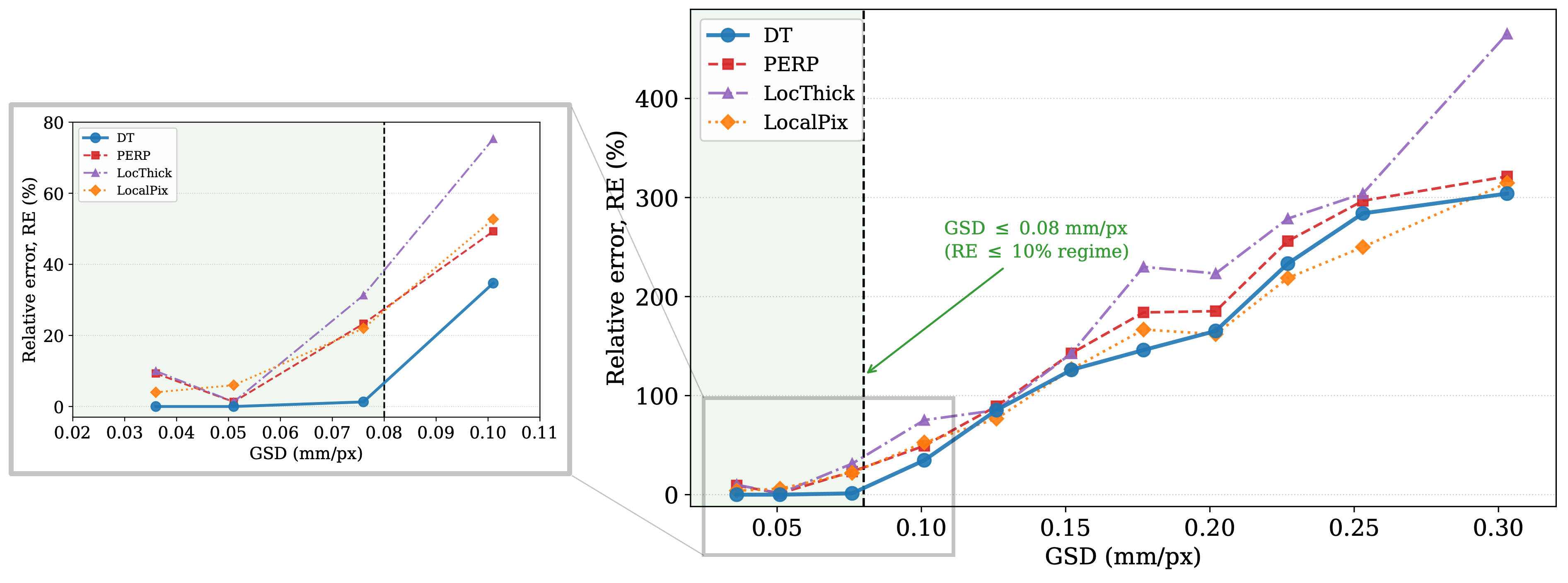}
    \caption{Comparison of crack-width estimation methods across GSD
             range. The distance-transform (DT) method provides the most accurate
             distance estimation of crack geometry, yielding the lowest mean relative
             error (125.5\%) and mean absolute error (1.88\,mm) in crack width measurement among the evaluated
             estimators.}
    \label{fig:width_estimators}
\end{figure}

\textbf{Recommended GSD.}
Table~\ref{table:width_methods_s15} and Fig.~\ref{fig:width_estimators} show that crack-width accuracy deteriorates as GSD increases. Specifically, at $\mathrm{GSD}\leq0.076$,mm/px ($d\leq30$,cm), DT estimates closely match the caliper reference, with RE below $1.3\%$. The error increases to $34.7\%$ at $0.101$,mm/px and reaches $304\%$ at $0.303$,mm/px, indicating substantial width overestimation at coarser resolutions. Increasing GSD also leads to more fragmented segmentation masks (Fig.~\ref{fig:sample_15_distance_exp}), where at $d=120$,cm, only part of the crack is recovered in four disconnected components. Nevertheless, the reference location remains measurable and follows the same GSD-dependent error trend.
The threshold-linear model in Fig.~\ref{fig:width_estimators} predicts $10\%$ RE at a GSD of $0.083$,mm/px, consistent with the empirical transition between $0.076$ and $0.101$,mm/px. Together, these results support a recommended GSD of $\mathbf{0.08}$,\textbf{mm/px or finer} for crack-width measurement with a target precision of $0.1$,mm. This criterion ensures that the scene resolution is finer than the target measurement precision. In practice, the required GSD can be achieved by adjusting stand-off distance and focal length using Eq.~\eqref{eq:gsd_iphone}, where a longer focal length permits a greater stand-off distance for the same GSD, whereas a shorter focal length requires closer imaging.

\subsection{End-to-End Crack Measurement Validation} \label{subsec:measurement_results}

\begin{figure}
    \centering
    \begin{minipage}[c]{0.06\linewidth}
        \centering
        \rotatebox[origin=c]{90}{\small\textbf{Crack 1}}
    \end{minipage}
    \begin{subfigure}[c]{0.92\linewidth}
        \raggedright
        \includegraphics[width=0.87\linewidth]{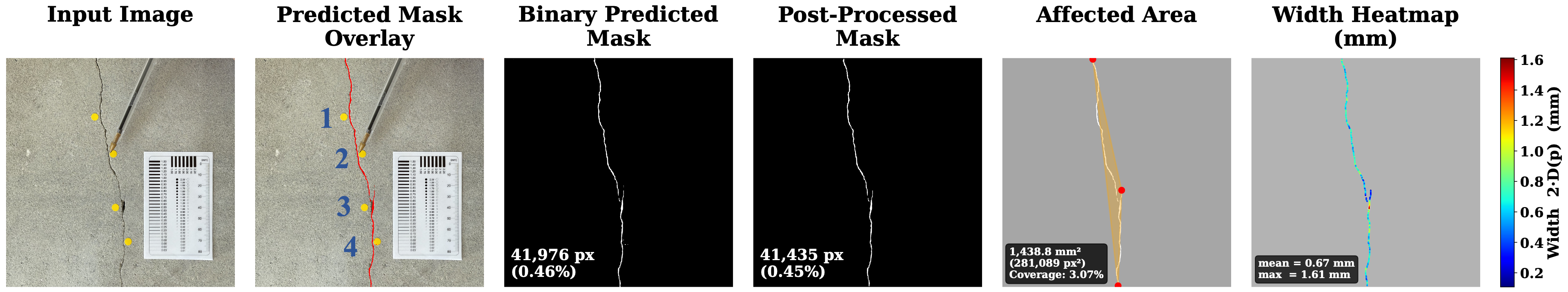}
        \label{fig:linear_crack_panel_1}
    \end{subfigure}

    \vspace{2pt}

    \begin{minipage}[c]{0.06\linewidth}
        \centering
        \rotatebox[origin=c]{90}{\small\textbf{Crack 2}}
    \end{minipage}
    \begin{subfigure}[c]{0.92\linewidth}
        \raggedright
        \includegraphics[width=0.87\linewidth]{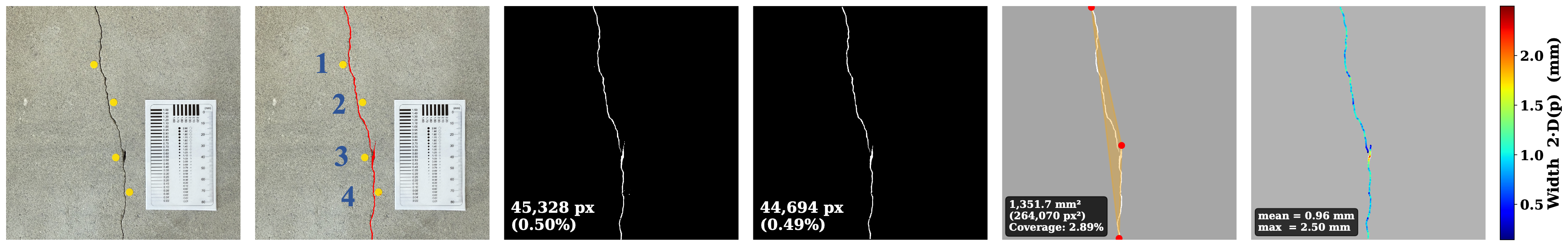}
        \label{fig:linear_crack_panel_2}
    \end{subfigure}

    \vspace{2pt}

    \begin{minipage}[c]{0.06\linewidth}
        \centering
        \rotatebox[origin=c]{90}{\small\textbf{Crack 3}}
    \end{minipage}
    \begin{subfigure}[c]{0.92\linewidth}
        \raggedright
        \includegraphics[width=0.87\linewidth]{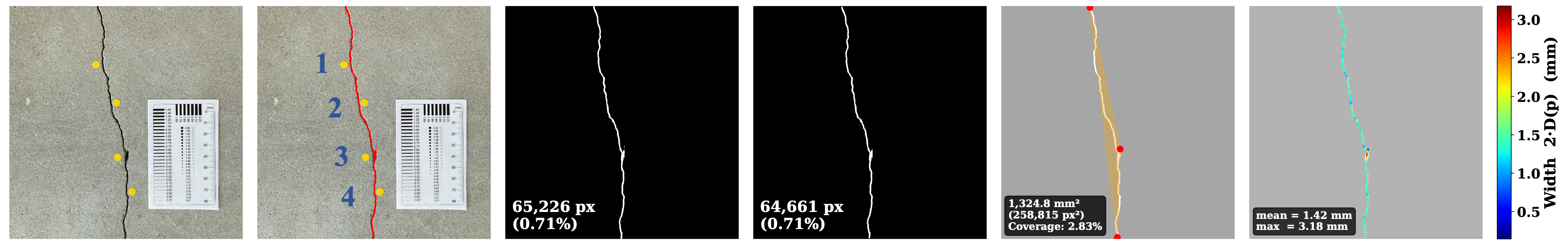}
        \label{fig:linear_crack_panel_3}
    \end{subfigure}

    \vspace{2pt}

    \begin{minipage}[c]{0.06\linewidth}
        \centering
        \rotatebox[origin=c]{90}{\small\textbf{Crack 4}}
    \end{minipage}
    \begin{subfigure}[c]{0.92\linewidth}
        \raggedright
        \includegraphics[width=0.87\linewidth]{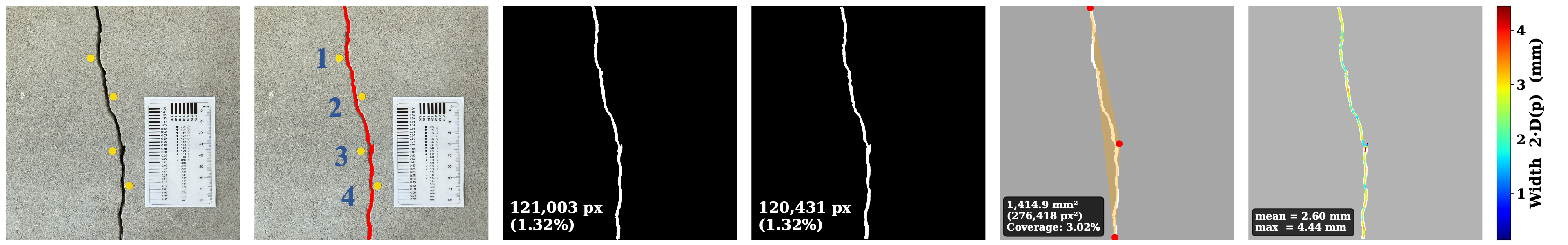}
        \label{fig:linear_crack_panel_4}
    \end{subfigure}

    \vspace{2pt}

    \begin{minipage}[c]{0.06\linewidth}
        \centering
        \rotatebox[origin=c]{90}{\small\textbf{Crack 5}}
    \end{minipage}
    \begin{subfigure}[c]{0.92\linewidth}
        \raggedright
        \includegraphics[width=0.87\linewidth]{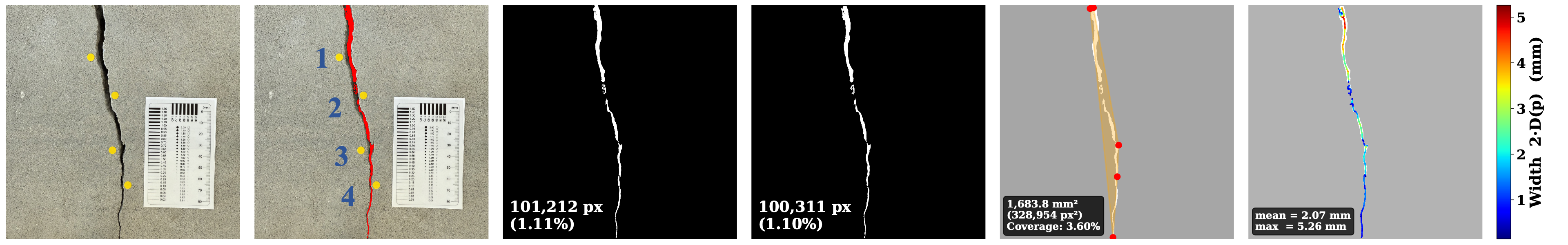}
        \label{fig:linear_crack_panel_5}
    \end{subfigure}

    \vspace{2pt}

    \begin{minipage}[c]{0.06\linewidth}
        \centering
        \rotatebox[origin=c]{90}{\small\textbf{Crack 6}}
    \end{minipage}
    \begin{subfigure}[c]{0.92\linewidth}
        \raggedright
        \includegraphics[width=0.87\linewidth]{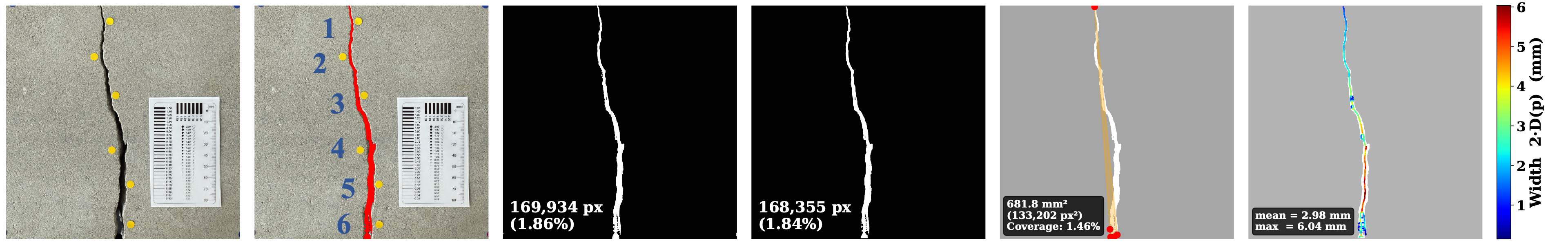}
        \label{fig:linear_crack_panel_6}
    \end{subfigure}

    \vspace{2pt}

    \begin{minipage}[c]{0.06\linewidth}
        \centering
        \rotatebox[origin=c]{90}{\small\textbf{Crack 7}}
    \end{minipage}
    \begin{subfigure}[c]{0.92\linewidth}
        \raggedright
        \includegraphics[width=0.87\linewidth]{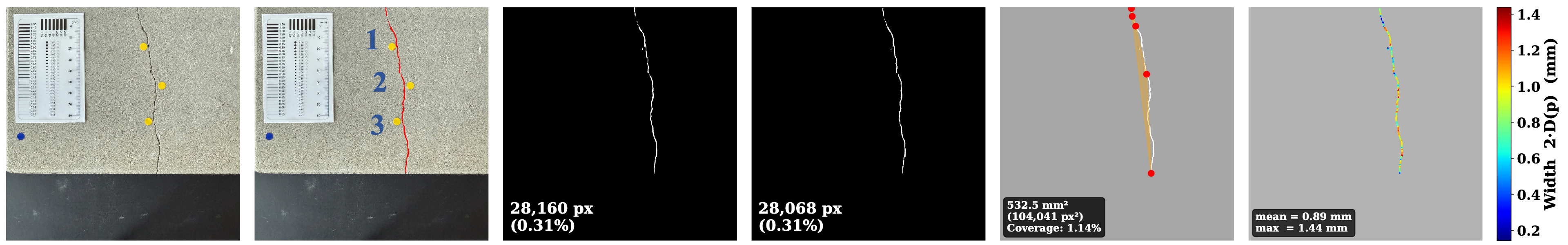}
        \label{fig:linear_crack_panel_7}
    \end{subfigure}

             \caption{End-to-end crack recognition outputs for the linear cracks.
             The numbers shown on the predicted-mask overlay identify the reference points
             at which the crack widths were measured (as given in Table \ref{tab:width_comparison_all}).}
    \label{fig:linear_cracks}
\end{figure}

\begin{figure}
    \centering
    \begin{minipage}[c]{0.06\linewidth}
        \centering
        \rotatebox[origin=c]{90}{\small\textbf{Web Crack 1}}
    \end{minipage}
    \begin{subfigure}[c]{0.92\linewidth}
        \raggedright
        \includegraphics[width=0.87\linewidth]{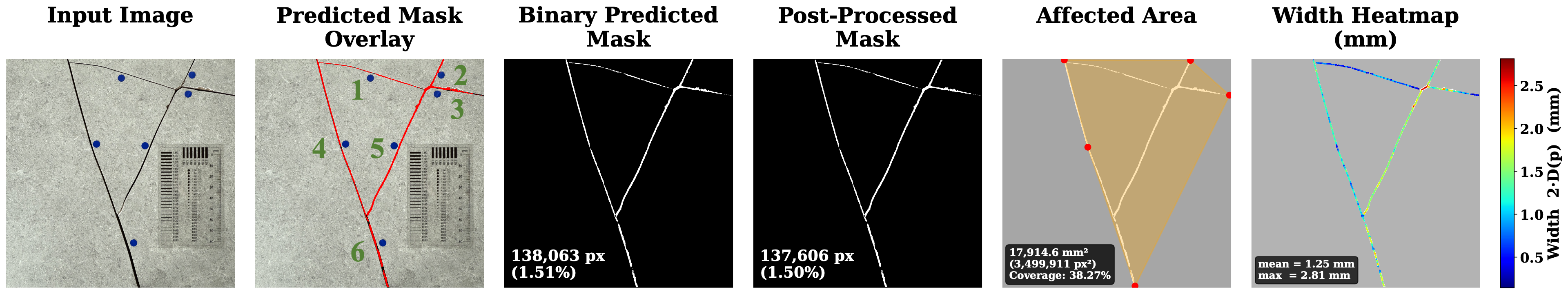}
        \label{fig:web_crack_panel_1}
    \end{subfigure}

    \vspace{2pt}

    \begin{minipage}[c]{0.06\linewidth}
        \centering
        \rotatebox[origin=c]{90}{\small\textbf{Web Crack 2}}
    \end{minipage}
    \begin{subfigure}[c]{0.92\linewidth}
        \raggedright
        \includegraphics[width=0.87\linewidth]{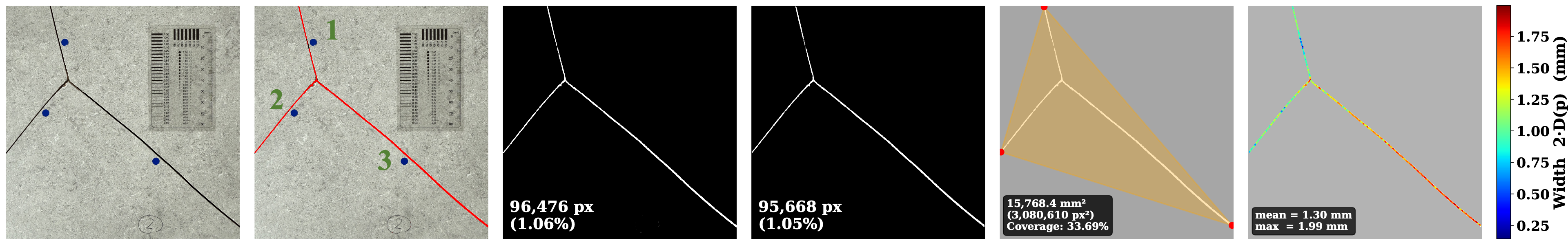}
        \label{fig:web_crack_panel_2}
    \end{subfigure}

    \vspace{2pt}

    \begin{minipage}[c]{0.06\linewidth}
        \centering
        \rotatebox[origin=c]{90}{\small\textbf{Web Crack 3}}
    \end{minipage}
    \begin{subfigure}[c]{0.92\linewidth}
        \raggedright
        \includegraphics[width=0.87\linewidth]{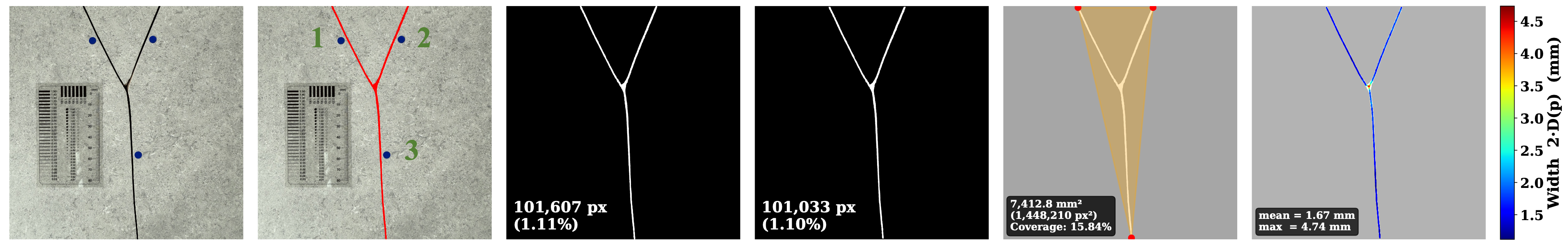}
        \label{fig:web_crack_panel_3}
    \end{subfigure}

    \caption{End-to-end crack recognition outputs for the web cracks.
             The numbers shown on the predicted-mask overlay identify the reference points
             at which the crack widths were measured (as given in Table \ref{tab:width_comparison_all}).}
    \label{fig:web_cracks}
\end{figure}

\begin{figure}
    \centering
    \begin{subfigure}[t]{0.32\linewidth}
        \centering
        \includegraphics[width=\linewidth]{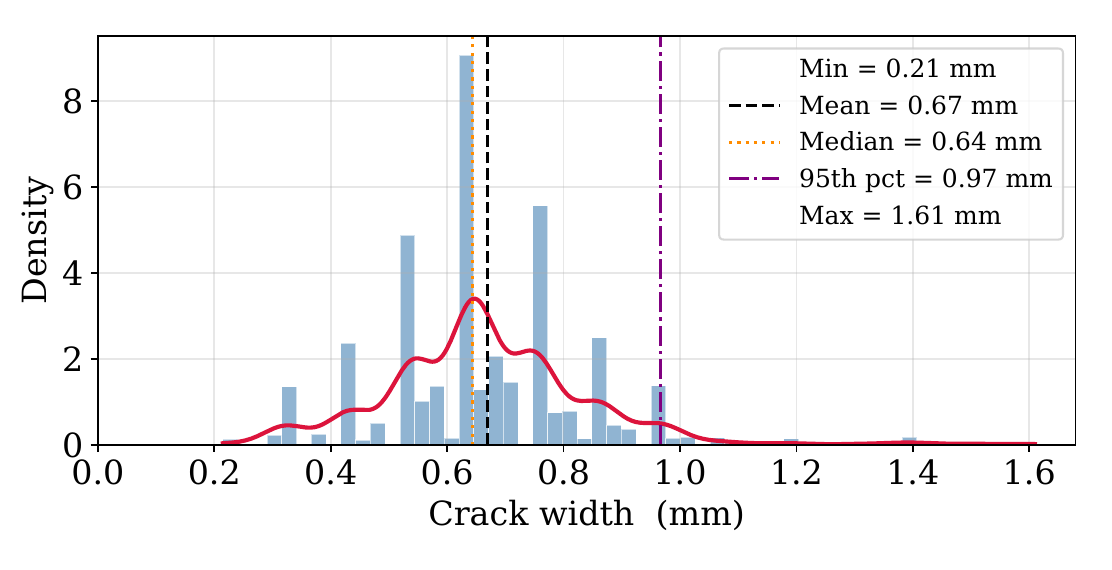}
        \caption{Crack 1}
        \label{fig:web_width_dist_1}
    \end{subfigure}
    \hfill
    \begin{subfigure}[t]{0.32\linewidth}
        \centering
        \includegraphics[width=\linewidth]{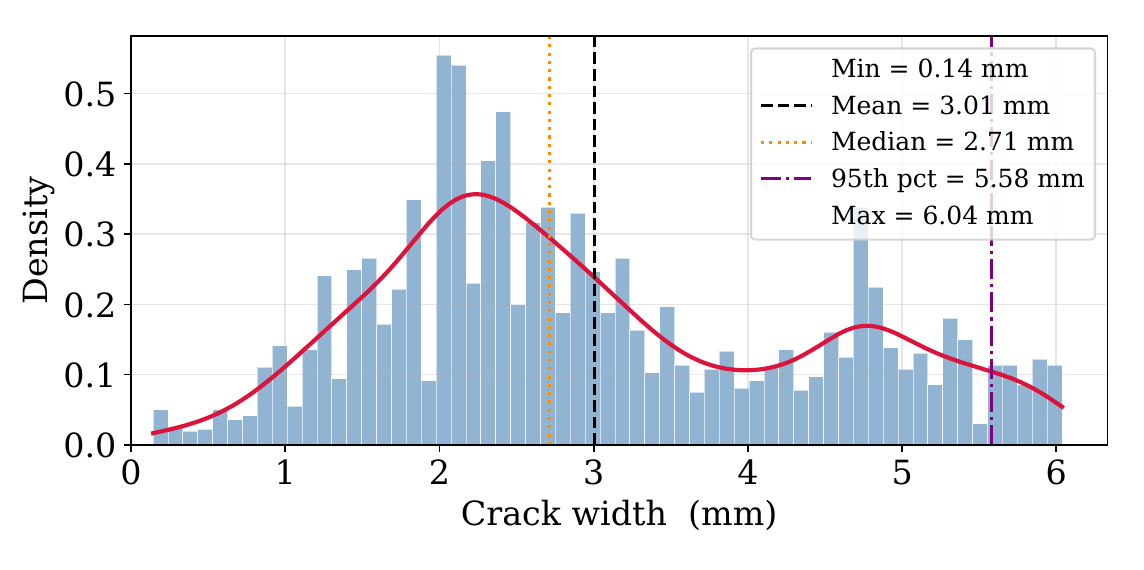}
        \caption{Crack 6}
        \label{fig:web_width_dist_2}
    \end{subfigure}
    \hfill
    \begin{subfigure}[t]{0.32\linewidth}
        \centering
        \includegraphics[width=\linewidth]{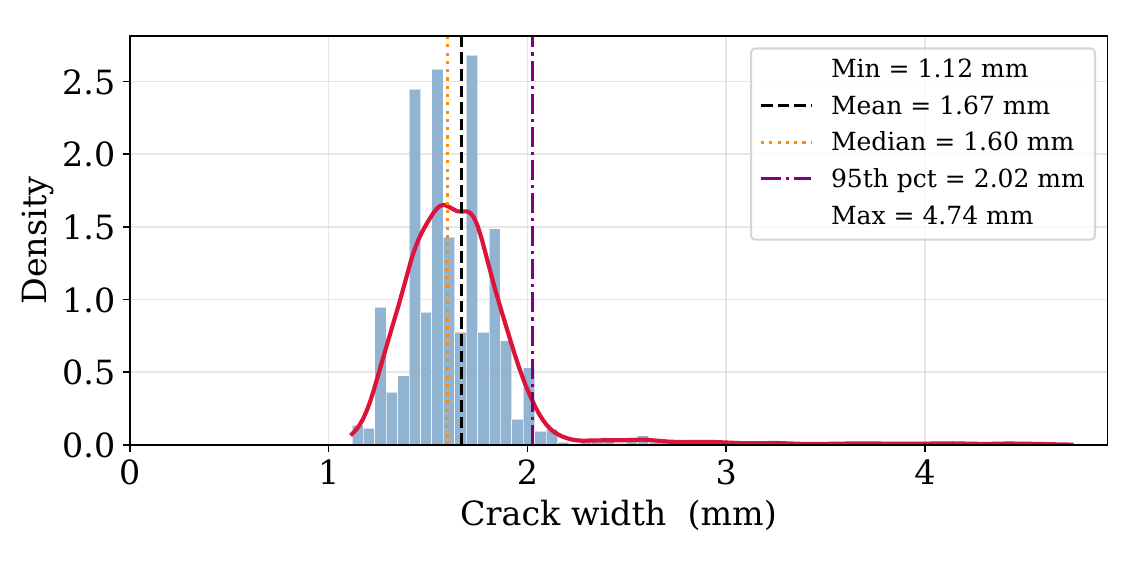}
        \caption{Web Crack 3}
        \label{fig:web_width_dist_3}
    \end{subfigure}
    \captionsetup{skip=0.5pt}
    \caption{Representative crack-width distributions for Crack 1, Crack 6, and Web Crack 3,
             obtained by applying kernel density estimation to the local distance-transform-based
             width measurements. The distributions characterize the variation in width along each
             predicted crack mask and enable the maximum crack width, $W_{\max}$, to be derived for
             each crack.}
    \label{fig:width_distributions}
\end{figure}

The end-to-end measurement accuracy was evaluated using images collected for this study from
seven linear and three web (branching) cracks, shown in panels (b) and (c) of
Figure~\ref{fig:sample_acquisition}. All images were acquired at a $20$\,cm stand-off distance using the
$1\times$ ($f_{35}=24$\,mm) setting of an iPhone~16 Pro Max with a 12\,MP macro-mode output
($N_{\mathrm{short}}=3024$\,px). Equation~\eqref{eq:gsd_iphone} gives a physics-based GSD of
$0.072$\,mm/px, while an independent ruler calibration gives $0.069$\,mm/px, a $3.7\%$
difference that confirms close agreement.
Across the ten cracks, 41 reference widths were measured with a digital caliper and compared
with estimates from the predicted masks. Table~\ref{tab:width_comparison_all} provides the
pointwise absolute and relative errors. At $\mathrm{GSD}=0.072$\,mm/px, the 29 linear-crack
locations yield an MRE of $6.62\%$, whereas the 12 geometrically complex web-crack locations
yield an MRE of $9.05\%$. The aggregate MRE is $7.33\%$. All three values remain below $10\%$, providing end-to-end validation within the recommended
$\mathrm{GSD}\leq0.08$\,mm/px range established in Section~\ref{subsec:gsd}.
The framework also derives geometric descriptors for severity analysis from each predicted mask.
The segmented crack area, $A_{\mathrm{crack}}$, is calculated from the foreground-pixel count, while
the affected area, $A$, is defined by the polygon enclosing the outermost crack endpoints.
Figures~\ref{fig:linear_cracks} and \ref{fig:web_cracks} present these areas with the segmentation
and width results for the linear and web cracks. Sample width distributions are shown in
Figure~\ref{fig:width_distributions}, where kernel density estimation of the local width estimates characterizes within-crack width variability. The minimum, mean, and maximum width ($W_{\max}$) were therefore extracted
from each crack's width distribution.

\begin{table*}[t]
\centering
\caption{Pointwise reference ($w^*$) and estimated ($\hat{w}$) crack widths at
         $\mathrm{GSD}=0.072$\,mm/px. C$x$-P$z$ denotes
         reference point $z$ on crack $x$, whereas W$x$-P$z$ denotes reference
         point $z$ on web crack $x$. AE and RE denote absolute and relative error.}
\label{tab:width_comparison_all}
\scriptsize
\setlength{\tabcolsep}{3pt}
\renewcommand{\arraystretch}{1.05}
\begin{tabular}{@{}c c c c c @{\hspace{8pt}} c c c c c@{}}
\toprule
Ref. Point & $w^*$ (mm) & $\hat{w}$ (mm) & AE (mm) & RE (\%) &
Ref. Point & $w^*$ (mm) & $\hat{w}$ (mm) & AE (mm) & RE (\%) \\
\midrule
C1-P1 & 0.50 & 0.58 & 0.08 & 16.00 & C6-P2 & 3.10 & 3.10 & 0.00 & 0.00 \\
C1-P2 & 0.50 & 0.58 & 0.08 & 16.00 & C6-P3 & 3.90 & 3.80 & 0.10 & 2.56 \\
C1-P3 & 1.70 & 1.61 & 0.09 & 5.29  & C6-P4 & 7.60 & 6.00 & 1.60 & 21.05 \\
C1-P4 & 0.50 & 0.55 & 0.05 & 10.00 & C6-P5 & 6.20 & 6.00 & 0.20 & 3.23 \\
C2-P1 & 0.70 & 0.65 & 0.05 & 7.14  & C6-P6 & 6.00 & 5.10 & 0.90 & 15.00 \\
C2-P2 & 0.70 & 0.70 & 0.00 & 0.00  & C7-P1 & 0.60 & 0.59 & 0.01 & 1.67 \\
C2-P3 & 2.70 & 2.50 & 0.20 & 7.41  & C7-P2 & 0.80 & 0.77 & 0.03 & 3.75 \\
C2-P4 & 0.60 & 0.70 & 0.10 & 16.67 & C7-P3 & 0.80 & 0.77 & 0.03 & 3.75 \\
C3-P1 & 1.20 & 1.20 & 0.00 & 0.00  & W1-P1 & 0.70 & 0.80 & 0.10 & 14.29 \\
C3-P2 & 1.20 & 1.10 & 0.10 & 8.33  & W1-P2 & 0.80 & 1.01 & 0.21 & 26.25 \\
C3-P3 & 3.20 & 3.18 & 0.02 & 0.63  & W1-P3 & 1.30 & 1.20 & 0.10 & 7.69 \\
C3-P4 & 1.10 & 1.15 & 0.05 & 4.55  & W1-P4 & 1.40 & 1.30 & 0.10 & 7.14 \\
C4-P1 & 3.10 & 3.00 & 0.10 & 3.23  & W1-P5 & 1.20 & 1.10 & 0.10 & 8.33 \\
C4-P2 & 3.10 & 3.00 & 0.10 & 3.23  & W1-P6 & 2.30 & 2.20 & 0.10 & 4.35 \\
C4-P3 & 4.90 & 4.44 & 0.46 & 9.39  & W2-P1 & 0.90 & 0.80 & 0.10 & 11.11 \\
C4-P4 & 2.90 & 3.01 & 0.11 & 3.79  & W2-P2 & 1.00 & 0.95 & 0.05 & 5.00 \\
C5-P1 & 4.10 & 4.40 & 0.30 & 7.32  & W2-P3 & 1.30 & 1.35 & 0.05 & 3.85 \\
C5-P2 & 4.00 & 3.90 & 0.10 & 2.50  & W3-P1 & 1.30 & 1.20 & 0.10 & 7.69 \\
C5-P3 & 4.40 & 4.00 & 0.40 & 9.09  & W3-P2 & 1.60 & 1.70 & 0.10 & 6.25 \\
C5-P4 & 1.90 & 1.80 & 0.10 & 5.26  & W3-P3 & 1.50 & 1.40 & 0.10 & 6.67 \\
C6-P1 & 1.90 & 2.00 & 0.10 & 5.26  &       &      &      &      &      \\
\midrule
\multicolumn{5}{l}{Linear Cracks 1-7: MAE $=0.19$\,mm; MRE $=6.62\%$} &
\multicolumn{5}{l}{Web cracks: MAE $=0.10$\,mm; MRE $=9.05\%$} \\
\bottomrule
\end{tabular}
\end{table*}

\subsection{Crack Severity Assessment} \label{subsec:severity_assessment}

\begin{figure}[h]
    \centering
    \includegraphics[width=0.8\textwidth]{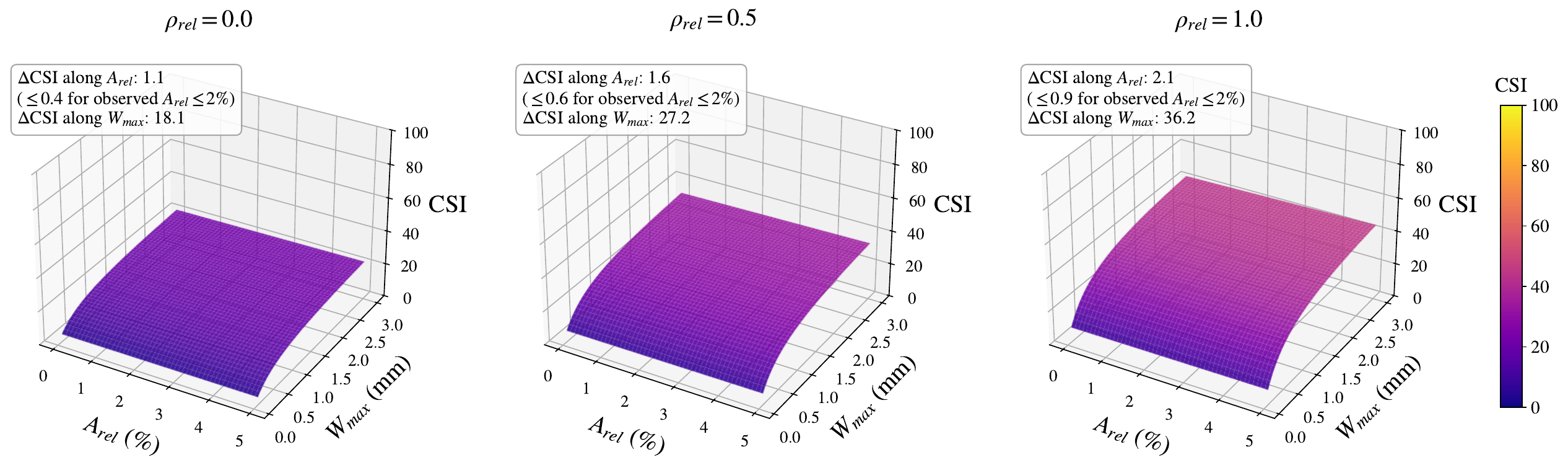}
    \caption{Ablation of the segmented crack-area term in the CSI
             (Eq.~\eqref{eq:csi_full}, $q=n=m=1$). Across the observed
             $A_{rel}\leq2\%$ range, changing $A_{rel}$ shifts the CSI by at most
             $0.4$-$0.9$ points, indicating negligible discriminatory contribution
             relative to $W_{\max}$.}
    \label{fig:csi_arel_ablation}
\end{figure}

The crack severity index (CSI) introduced in Equation \ref{eq:csi_full} initially combines maximum
crack width ($W_{\max}$), affected-area ratio ($\rho_{rel}$), and segmented crack-area
ratio ($A_{rel}$). 
Further analysis showed that $A_{rel}$ provided negligible discriminatory information to the CSI. The computed values of $A_{rel}$ for the ten measured cracks were approximately within $0$-$2\%$, over which the unpowered factor $(1+A_{rel})/2$ varies only from $0.50$ to $0.51$. This is due to the fact that crack pixels comprise a small portion of the image. In OmniCrack30k, crack pixels average only 1.9\% per image in the dataset.
Consequently, factor ablations showed this term changes the CSI by no more than $0.9$ points over the observed range, whereas variation in
$W_{\max}$ changes it by $18$-$36$ points
(Figure~\ref{fig:csi_arel_ablation}).
In addition, $A_{rel}$ partially overlaps the spatial
information represented by $\rho_{rel}$. Therefore, the segmented-area term was removed
from the CSI. 
The resulting two-factor CSI is:
\vspace{-12pt}
\begin{equation} \label{eq:csi_reduced}
\mathrm{CSI}=100
\left[\frac{W_{\max}}{W_{\max}+W_{50}}\right]^{q}
\left[\frac{1+\rho_{rel}}{2}\right]^{n},
\end{equation}

\begin{figure}[h]
    \centering
    \includegraphics[width=\textwidth]{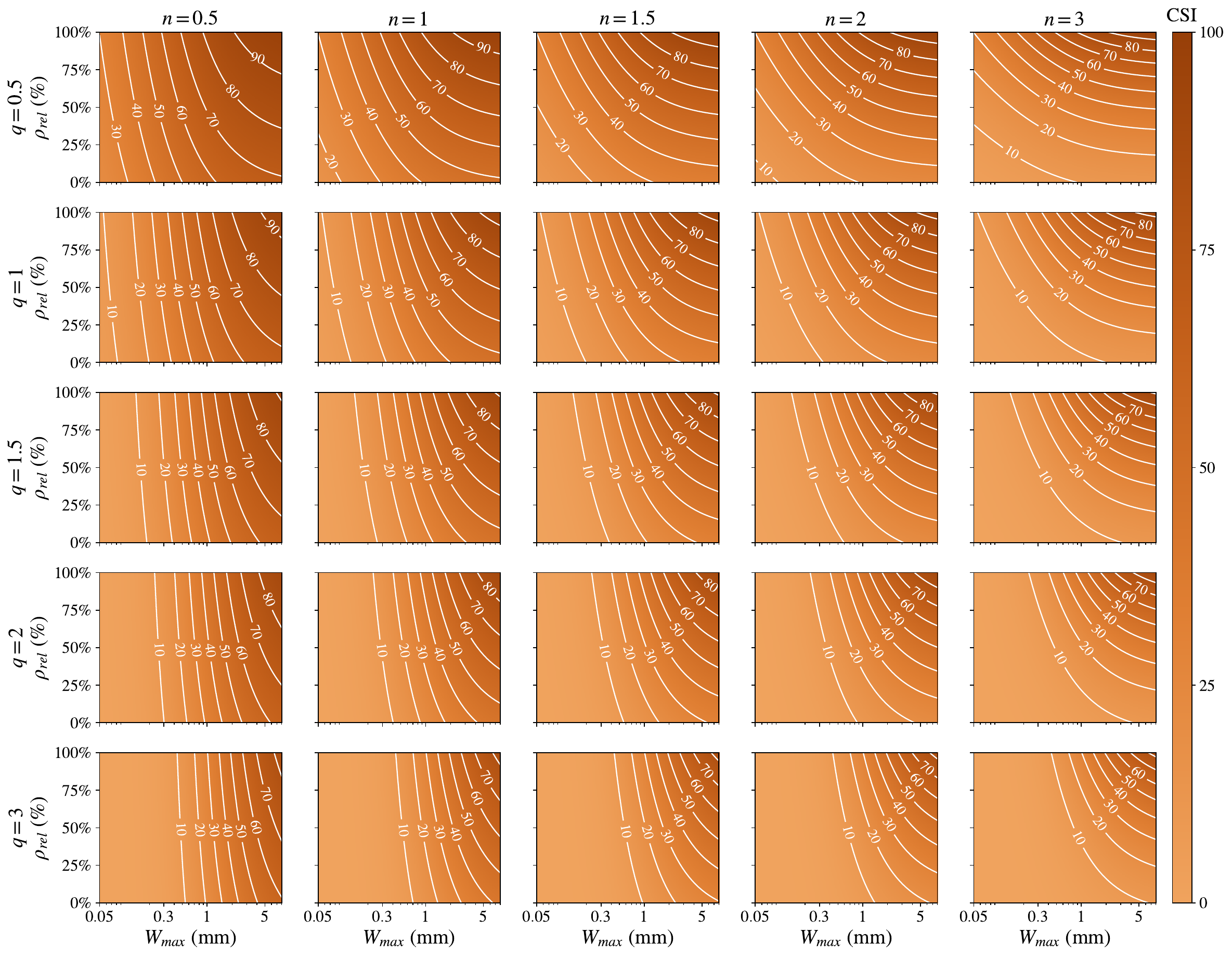}
    \caption{Sensitivity analysis of the CSI (Eq. \ref{eq:csi_reduced}) to the width exponent $q$ (rows) and
             affected-area exponent $n$ (columns), with $W_{50}=0.5$\,mm. Labeled
             iso-CSI contours show the trade-off between $W_{\max}$ and
             $\rho_{rel}$.}
    \label{fig:csi_power_ablation}
\end{figure}

\begin{figure}[h]
    \centering
    \includegraphics[width=0.8\textwidth]{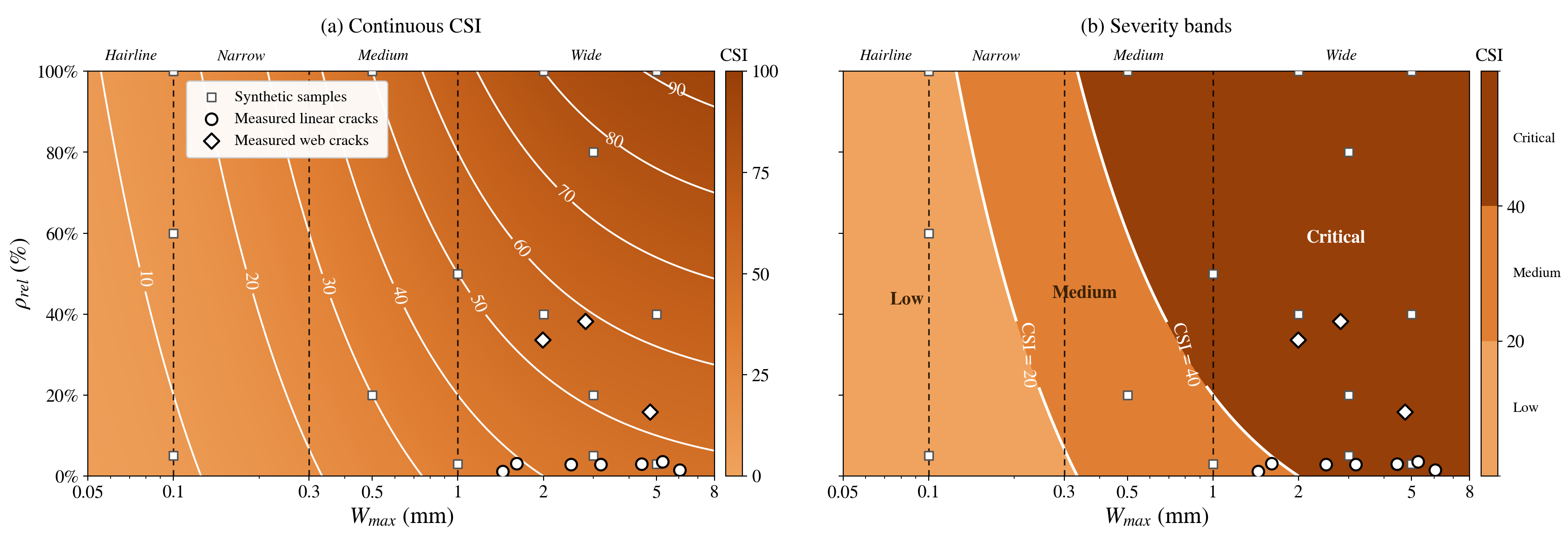}
    \caption{CSI severity map for an OSIM-aligned calibration
             ($q=n=1$, $W_{50}=0.5$\,mm): (a) continuous CSI contours and OSIM width boundaries depicted as dashed lines, 
             (b) Low, Medium, and Critical regions separated at recommended CSI values of 20 and
             40. Markers represent the
             synthetic and measured crack samples.}
    \label{fig:csi_severity_map}
\end{figure}

A sensitivity analysis varied $q$ and $n$ over $\{0.5,1,1.5,2,3\}$ to evaluate the relative
influence of crack width and affected extent. The result is given in Figure~\ref{fig:csi_power_ablation}. Values of
$q<1$ prematurely saturate the width response, whereas $q>1$ suppresses the response to narrow
and moderate cracks, and $n$ controls the influence of spatial extent. For inspection scenarios requiring consistency with OSIM width classes, the sensitivity maps support $(q,n)=(1,1)$ as a suitable calibration to be examined. Under this calibration, $W_{\max}$ remains the main severity measure, while
$\rho_{rel}$ increases the score for spatially distributed cracking. As shown in Figure~\ref{fig:csi_severity_map}, CSI thresholds of 20 and
40, for instance, can define \emph{Low} ($\mathrm{CSI}\leq20$), \emph{Medium} ($20<\mathrm{CSI}\leq40$), and
\emph{Critical} ($\mathrm{CSI}>40$) severity. At negligible
affected extent, these thresholds correspond to $W_{\max}\approx0.33$ and $2.0$\,mm,
respectively, consistent with the OSIM width classes.

Evaluation of 15 synthetic crack width and affected area combinations also confirmed that CSI increases monotonically
with both inputs and follows engineering intuition (Figure~\ref{fig:csi_severity_map}). Hairline cracks remain \emph{Low}, whereas a wide
opening or a moderate opening with large affected extent can reach the \emph{Critical} region.
Table~\ref{tab:csi_results_v2} provides results for the ten measured cracks. Five wide linear cracks were
classified as \emph{Critical} and the two narrowest as \emph{Medium}. All three web cracks were
\emph{Critical} and received the highest scores ($52.4$-$58.7$) due to their larger
affected-area ratios. By ranking branching networks above isolated cracks of comparable width,
CSI captures the contribution of affected extent. These thresholds apply to the present concrete
inspection scenario. As $W_{50}$, $q$, $n$, and the decision thresholds are configurable,
CSI can be recalibrated for other inspection regulations, including pavement-specific manuals.

\begin{table}[h]
\small
\renewcommand{\arraystretch}{1.05}
\centering
\caption{CSI assessment of the ten measured linear and web cracks using
         $q=n=1$, $W_{50}=0.5$\,mm, and severity thresholds of 20 and 40.}
\label{tab:csi_results_v2}
\begin{tabular}{l c c c c c c}
    \hline\hline
    Sample & $W_{\max}$ (mm) & $\rho_{rel}$ (\%) & Width term & Extent term & CSI & Severity \\
    \hline
    Crack 1     & 1.61 & 3.07  & 0.76 & 0.52 & 39.3 & Medium \\
    Crack 2     & 2.50 & 2.89  & 0.83 & 0.51 & 42.9 & Critical \\
    Crack 3     & 3.18 & 2.83  & 0.86 & 0.51 & 44.4 & Critical \\
    Crack 4     & 4.44 & 3.02  & 0.90 & 0.52 & 46.3 & Critical \\
    Crack 5     & 5.26 & 3.60  & 0.91 & 0.52 & 47.3 & Critical \\
    Crack 6     & 6.04 & 1.46  & 0.92 & 0.51 & 46.9 & Critical \\
    Crack 7     & 1.44 & 1.14  & 0.74 & 0.51 & 37.5 & Medium \\
    Web Crack 1 & 2.81 & 38.27 & 0.85 & 0.69 & 58.7 & Critical \\
    Web Crack 2 & 1.99 & 33.69 & 0.80 & 0.67 & 53.4 & Critical \\
    Web Crack 3 & 4.74 & 15.84 & 0.90 & 0.58 & 52.4 & Critical \\
    \hline\hline
\end{tabular}
\end{table}

\section{Conclusion and Future Work} \label{sec:conclusion}

This study proposes Inspect Any Crack (IAC), an end-to-end crack recognition and condition assessment framework integrating crack segmentation, calibrated geometric measurement, and continuous severity assessment. SACv2-Tiny was developed through efficient adaptation and systematic evaluation of SAM2 variants for accurate, robust, and efficient crack segmentation. The model achieved a $62.49\%$ F1-score, $45.45\%$ IoU on OmniCrack30K, and $9.94$~FPS while updating only $29.8$K parameters, and demonstrated strong zero-shot robustness across three unseen datasets. For physical crack measurement, comparison of four width estimators identified the distance-transform method as the most accurate, while resolution analysis established a recommended ground sampling distance (GSD) of $0.08$~mm/px or finer for reliable measurement. End-to-end validation at 41 caliper-referenced locations yielded an aggregate MRE of $7.33\%$. Furthermore, an innovative Crack Severity Index (CSI) was proposed, which combines maximum crack width and affected-area ratio into a bounded, configurable severity score adaptable to inspection standards.
Currently, the CSI considers only crack width and spatial extent. Future work can incorporate additional characteristics, including crack orientation and pattern type, into severity assessment.

\section{Data Availability Statement}
Some or all data, models, or code that support the findings of this study are available from the corresponding author upon reasonable request.

\section{Acknowledgment}
This research was funded by the Discovery Grant (Grant No.: RGPIN-2022-05151) of the
Natural Sciences and Engineering Research Council of Canada (NSERC).

\clearpage
\begingroup
\setstretch{1.0}
\bibliography{references}
\endgroup

\end{document}